\documentclass[sigconf]{acmart}
\AtBeginDocument{%
  \providecommand\BibTeX{{%
    \normalfont B\kern-0.5em{\scshape i\kern-0.25em b}\kern-0.8em\TeX}}}

\setcopyright{acmlicensed}
\copyrightyear{2025}
\acmYear{2025}
\acmDOI{XXXXXXX.XXXXXXX}

\acmConference[KDD '25]{the 31th ACM SIGKDD Conference on Knowledge Discovery and Data Mining }{August 03 – 07, 2025}{Toronto, ON, Canada}
\acmISBN{978-1-4503-XXXX-X/18/06}

\usepackage[ruled,vlined,linesnumbered]{algorithm2e}
\usepackage{enumitem}
\usepackage{multirow}
\usepackage[normalem]{ulem}
\usepackage{subcaption}

\usepackage[page]{appendix} % print appendices title
\usepackage{lipsum}

\begin{document}

%%
%% The "title" command has an optional parameter,
%% allowing the author to define a "short title" to be used in page headers.
\title{Controlling Diversity at Inference: Guiding Diffusion \\ Recommender Models with Targeted Category Preferences}

%%
%% The "author" command and its associated commands are used to define
%% the authors and their affiliations.
%% Of note is the shared affiliation of the first two authors, and the
%% "authornote" and "authornotemark" commands
%% used to denote shared contribution to the research.

\author{Gwangseok Han}
\affiliation{
    \institution{Pohang University of \\ Science and Technology}
    % \city{Pohang}
    % \state{Gyeongbuk}
    \country{Pohang, Republic of Korea}
}
\authornote{Both authors contributed equally to this work}
\email{gshan@postech.ac.kr}

\author{Wonbin Kweon}
\affiliation{
    \institution{Pohang University of \\ Science and Technology}
    % \city{Pohang}
    % \state{Gyeongbuk}
    \country{Pohang, Republic of Korea}
}
\authornotemark[1]
\email{kwb4453@postech.ac.kr}

\author{Minsoo Kim}
\affiliation{
    \institution{Pohang University of \\ Science and Technology}
    % \city{Pohang}
    % \state{Gyeongbuk}
    \country{Pohang, Republic of Korea}
}
\email{km19809@postech.ac.kr}

\author{Hwanjo Yu}
\affiliation{
    \institution{Pohang University of \\ Science and Technology}
    % \city{Pohang}
    % \state{Gyeongbuk}
    \country{Pohang, Republic of Korea}
}
\authornote{Corresponding author.}
\email{hwanjoyu@postech.ac.kr}

%%
%% By default, the full list of authors will be used in the page
%% headers. Often, this list is too long, and will overlap
%% other information printed in the page headers. This command allows
%% the author to define a more concise list
%% of authors' names for this purpose.
\renewcommand{\shortauthors}{Gwangseok Han*, Wonbin Kweon*, Minsoo Kim, and Hwanjo Yu}

\setlength\abovecaptionskip{+0.1pt}

%%
%% The abstract is a short summary of the work to be presented in the
%% article.
\begin{abstract} 

Diversity control is an important task to alleviate bias amplification and filter bubble problems.
The desired degree of diversity may fluctuate based on users' daily moods or business strategies.
However, existing methods for controlling diversity often lack flexibility, as diversity is decided during training and cannot be easily modified during inference.
We propose \textbf{D3Rec} (\underline{D}isentangled \underline{D}iffusion model for \underline{D}iversified \underline{Rec}ommendation), an end-to-end method that controls the accuracy-diversity trade-off at inference.
D3Rec meets our three desiderata by (1) generating recommendations based on category preferences, (2) controlling category preferences during the inference phase, and (3) adapting to arbitrary targeted category preferences.
In the forward process, D3Rec removes category preferences lurking in user interactions by adding noises.
Then, in the reverse process, D3Rec generates recommendations through denoising steps while reflecting desired category preferences.
Extensive experiments on real-world and synthetic datasets validate the effectiveness of D3Rec in controlling diversity at inference. 
\end{abstract}

%%
%% The code below is generated by the tool at http://dl.acm.org/ccs.cfm.
%% Please copy and paste the code instead of the example below.
%%

\begin{CCSXML}
<ccs2012>
<concept>
<concept_id>10002951.10003317.10003347.10003350</concept_id>
<concept_desc>Information systems~Recommender systems</concept_desc>
<concept_significance>500</concept_significance>
</concept>
</ccs2012>
\end{CCSXML}

\ccsdesc[500]{Information systems~Recommender systems}

% \begin{CCSXML}
% <ccs2012>
%  <concept>
%   <concept_id>00000000.0000000.0000000</concept_id>
%   <concept_desc>Do Not Use This Code, Generate the Correct Terms for Your Paper</concept_desc>
%   <concept_significance>500</concept_significance>
%  </concept>
%  <concept>
%   <concept_id>00000000.00000000.00000000</concept_id>
%   <concept_desc>Do Not Use This Code, Generate the Correct Terms for Your Paper</concept_desc>
%   <concept_significance>300</concept_significance>
%  </concept>
%  <concept>
%   <concept_id>00000000.00000000.00000000</concept_id>
%   <concept_desc>Do Not Use This Code, Generate the Correct Terms for Your Paper</concept_desc>
%   <concept_significance>100</concept_significance>
%  </concept>
%  <concept>
%   <concept_id>00000000.00000000.00000000</concept_id>
%   <concept_desc>Do Not Use This Code, Generate the Correct Terms for Your Paper</concept_desc>
%   <concept_significance>100</concept_significance>
%  </concept>
% </ccs2012>
% \end{CCSXML}

% \ccsdesc[500]{Do Not Use This Code~Generate the Correct Terms for Your Paper}
% \ccsdesc[300]{Do Not Use This Code~Generate the Correct Terms for Your Paper}
% \ccsdesc{Do Not Use This Code~Generate the Correct Terms for Your Paper}
% \ccsdesc[100]{Do Not Use This Code~Generate the Correct Terms for Your Paper}

%%
%% Keywords. The author(s) should pick words that accurately describe
%% the work being presented. Separate the keywords with commas.
\keywords{Diversity, User-controllable recommendation systems, Generative Recommendation}

%% A "teaser" image appears between the author and affiliation
%% information and the body of the document, and typically spans the
%% page.
% \begin{teaserfigure}
%   \includegraphics[width=\textwidth]{sampleteaser}
%   \caption{Seattle Mariners at Spring Training, 2010.}
%   \Description{Enjoying the baseball game from the third-base
%   seats. Ichiro Suzuki preparing to bat.}
%   \label{fig:teaser}
% \end{teaserfigure}

% \received{20 February 2007}
% \received[revised]{12 March 2009}
% \received[accepted]{5 June 2009}

%%
%% This command processes the author and affiliation and title
%% information and builds the first part of the formatted document.
\maketitle

\section{Introduction}
In the information-overloaded era, recommendation systems are crucial for online services, helping to filter out irrelevant items for users \cite{DiffRec, COR, kweon2024top, lee2024continual}.
Recommender systems infer user preferences based on past behaviors (e.g., clicks) and item characteristics (e.g., category), and generate recommendation lists tailored to users' tastes.
However, as systems are focused on accuracy, bias amplification problem \cite{Calib, DecRs, kweon2024doubly} is raised as a critical issue.
Due to the uneven distribution of item categories (e.g., movie genres) in users' historical data, recommender systems often favor popular items and categories, which can narrow user interests and create a filter bubble \cite{DecRs}.
Addressing the filter bubble is essential, as it can negatively impact user satisfaction and reduce the uniqueness of items in the long term \cite{UCRS}.
Consequently, there is growing interest in controlling the diversity of recommendations \cite{ctrl_fible, UCRS, DGRec}.
Some studies \cite{FDSB, DCRS} have shown through online A/B tests that increasing the diversity of recommended items can lead to higher user engagement.

\begin{figure}[t]
  \centering
  \includegraphics[width=\linewidth]{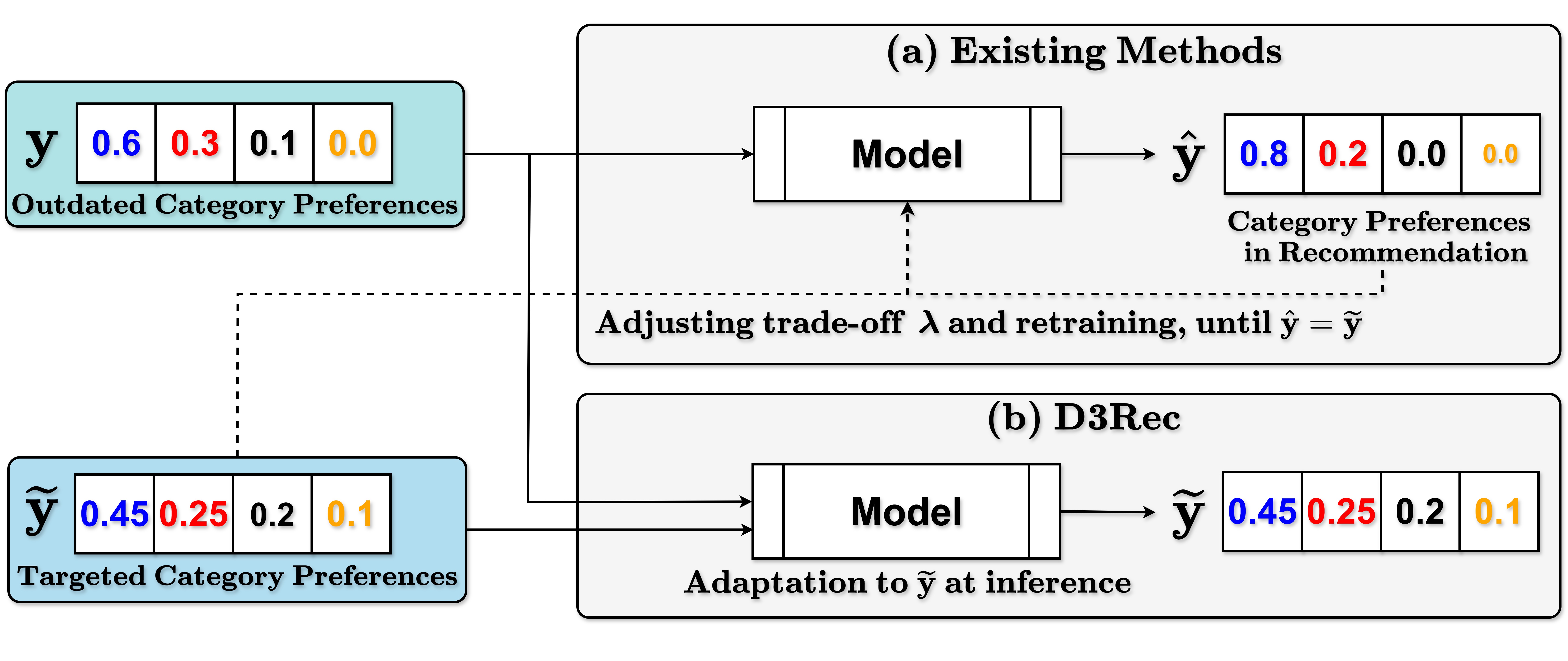}
  \caption{Diversity control: (a) Existing methods \cite{DCRS, DGCN} and (b) D3Rec (ours).} % $\boldsymbol{\lambda}$ is a hyper-parameter fixed in the training phase.
  \label{fig:diversity_process}
  \vspace{-0.5cm}
\end{figure}

Existing methods for controlling recommendation diversity can be broadly categorized into two types: post-processing and end-to-end approaches.
In the early stage, the post-processing modules \cite{div_first, dpp, pmf} are introduced to control diversity after generating candidate items.
However, since the post-processing modules operate independently of the candidate generation process, diversity signals are not incorporated during training, leading to suboptimal solutions \cite{DGCN, dilemma}.
To address this issue, end-to-end approaches \cite{DGCN, DCRS, DPCML} have been developed to jointly optimize accuracy and diversity during training by incorporating item category information.
While effective, these methods often lose the flexibility to adjust diversity on the fly.
The accuracy-diversity trade-off is decided and fixed by hyperparameters set during training and cannot be easily modified during inference.
For example, as illustrated in Figure \ref{fig:diversity_process} (a), if users want more or less diverse recommendations, they must consistently provide feedback (e.g., clicks and ratings) until the system adapts to their preferences \cite{UCRS}.
Alternatively, recommender models need to be re-trained after adjusting the hyperparameters controlling the accuracy-diversity trade-off.
These processes are inefficient and inadequate for adapting to targeted diversity, which may fluctuate based on users' daily moods or business strategies.

In this paper, we propose \textbf{D3Rec} (\underline{D}isentangled \underline{D}iffusion model for \underline{D}iversified \underline{Rec}ommendation), an end-to-end method that controls the accuracy-diversity trade-off \textit{at the inference phase}.
D3Rec is designed based on our three desiderata:
\vspace{-0.1cm}
\begin{enumerate}[leftmargin=*]
    \item \textbf{Generating} recommendations based on category preferences.
    \item \textbf{Controlling} category preferences during the inference phase.
    \item \textbf{Adapting} to arbitrary targeted category preferences.
\end{enumerate}
% \vspace{-0.2cm}
To achieve the above goals, we adopt a diffusion framework \cite{DDPM, classifierfree} to generate recommendations.
In the forward process, D3Rec eliminates category preferences lurking within user interactions by adding noises.
Then, in the reverse process, D3Rec generates recommendations through denoising steps while reflecting the targeted category preference.
Specifically, for the first desideratum, we employ \textit{disentangled two-tower encoders} to capture the category preferences.
For the second, we provide the reverse process with guidance toward targeted category preferences as a condition for the generation.
Lastly, we devise two auxiliary tasks to ensure that the generated recommendations align with the targeted category preferences.
These functionalities enable systems to control the diversity on the fly by flexibly changing desired category preferences, as depicted in Figure \ref{fig:diversity_process} (b).

D3Rec is highly applicable to various real-world scenarios involving different desired category distributions.
When users require diverse recommendations, the system can enhance the diversity by adjusting the targeted category preference smoothly.
On the other hand, the system can generate kindred and personalized recommendations by skewing the distribution of category preferences.
Moreover, the system can set arbitrary targeted preferences that may deviate from the original preferences.
This approach allows the system to adapt readily to the users' capricious moods (e.g., a user who typically prefers action movies might request romantic movies when with a partner).
To summarize, our main contributions are as follows:
% \vspace{-0.2cm}
\begin{itemize}[leftmargin=*]
  \item We propose three desiderata for methods that enhance recommendation diversity, enabling flexible adaptation to desired diversity at inference.
  \item We introduce D3Rec, which removes category preferences embedded in user interactions by adding noise and generates recommendations through denoising steps that reflect the targeted category preferences.
  \item We validate the superiority of D3Rec through extensive experiments on three real-world and synthetic datasets. We also provide in-depth analyses to verify the effectiveness of each proposed component.
\end{itemize}

\section{Related Work}
% \begin{figure}[t]
%   \centering
%   \includegraphics[width=\linewidth]{figure/causal.jpg}
%   \vspace{+0.005cm}
%   \caption{Causal graph of the generation process}
%   \label{fig:causal}
%   \vspace{-0.5cm}
% \end{figure}

\vspace{+0.1cm}
\textbf{Diversity Control in Recommendation Systems.}
Ziegler et al. \cite{div_first} first introduced a greedy algorithm \cite{MMR} as a post-processing module to balance accuracy and diversity.
Subsequently, several post-processing approaches \cite{dpp, ComiRec, FDSB, div_er, pmf} were developed to impose diversity using various measures, such as the determinantal point process \cite{dpp} and category distribution \cite{ComiRec}.
However, since these modules operate independently of the candidate generation process, diversity signals are not integrated during training, resulting in suboptimal solutions \cite{DGCN, dilemma}.
Also, post-processing approaches exhibit large inference latency due to their additional optimizations.
To address this issue, Zheng et al. \cite{DGCN} proposed an end-to-end approach.
Most end-to-end methods \cite{DGRec, DCRS, DPGNN, DPCML} leverage category information for different purposes, including sampling techniques \cite{DGCN}, re-weighting strategies based on global category popularity \cite{DGRec}, and classifying item categories to predict user preferences \cite{DCRS}.
While effective, these methods often lack the flexibility to adjust diversity dynamically.
The accuracy-diversity trade-off is set by hyperparameters during training and cannot be easily modified during inference.

\vspace{+0.1cm}
\noindent \textbf{Generative recommendation.}
Generative models are widely adopted across various domains \cite{multvae, stable, Diffusion_LM} due to their ability to model complex distributions.
In recommendation systems, these models are employed to capture users' non-linear and intricate preferences.
These methods can be broadly categorized into three groups: VAE-based methods \cite{multvae, macrid, COR}, Generative Adversarial Network (GAN)-based methods \cite{pd_gan, gan_rec}, and diffusion-based methods \cite{DiffRec, dreamrec}.
VAE-based methods utilize an encoder to approximate the posterior distribution and a decoder to estimate the probability of user interactions with non-interacted items.
GAN-based methods predict user interactions by using a generator, which is optimized through adversarial learning with the discriminator.
Recently, diffusion models have gained popularity for addressing the limitations of VAEs and GANs, such as posterior collapse and mode collapse.
In diffusion-based approaches, user interactions \cite{DiffRec} or target item representations \cite{dreamrec} are corrupted by progressively adding noise during the forward process.
Recommendations are then generated by denoising the corrupted data during the reverse process.

\section{Preliminary}
% In this section, we briefly review the basic diffusion model for recommendation systems, which is described in the \cite{DiffRec}. The diffusion process works in two steps: the forward (diffusion) process and the reverse process.

% \vspace{+0.15cm}
\noindent \textbf{Notations.} \label{sec:notation}
Let $\mathcal{U}$ be the set of users, $\mathcal{I}$ be the set of items, and $\mathcal{C}$ be the set of item categories in the dataset.
An item $i \in \mathcal{I}$ is associated with one or more categories and $\boldsymbol{F} \in \mathbb{R}^{|\mathcal{I}| \times |\mathcal{C}|}$ denotes the item-category matrix.
Each row $\boldsymbol{F}[i]$ denotes the category distribution of item $i$.
For example, if an item $i$ is associated with the second and the fourth categories among four categories, the category distribution would be represented as $\boldsymbol{F}[i] = [0, 0.5, 0, 0.5]^\top$.
The interaction history of a user $u \in \mathcal{U}$ is represented by $\boldsymbol{x}^u \in \{0, 1\}^{|\mathcal{I}|}$, where $x^u_i=1$ indicates that user $u$ interacted with item $i$, and 0 otherwise.
The category preference of a user $u$ is defined as $\boldsymbol{y}^u \in \mathbb{R}^{|\mathcal{C}|}$, by aggregating the category distributions of items with which the user has interacted: $\boldsymbol{y}^u = \boldsymbol{F}^\top\boldsymbol{x}^u / || \boldsymbol{F}^\top \boldsymbol{x}^u||_1$.

% $\boldsymbol{y}^u[j] = (\boldsymbol{F}^\top \cdot \boldsymbol{x}^u)[j] / \sum_i(\boldsymbol{F}^\top \cdot \boldsymbol{x}^u)[i]$.

% For example, let's contemplate four categories: $\{ c_1, c_2, c_3, c_4 \}$.
% If an item $i$ is associated with categories $c_2$ and $c_4$, the categories of item $i$ are represented as a vector $f_i$: $[0, 0.5, 0, 0.5]^\top$.
% Suppose user $u$ consumes items $i_1$ and $i_2$, where the related categories are $\{ c_2, c_4 \}$ and $\{ c_3 \}$ respectively.
% Then $\boldsymbol{y}^u = (f_{i_1} \oplus f_{i_2}) / \sum_k (f_{i_1} \oplus f_{i_2})_k = [0, 0.25, 0.5, 0.25]^\top$ where $\oplus$ denotes element-wise addition.

\vspace{+0.15cm}
\noindent \textbf{Diffusion model \cite{DDPM}.}  \label{sec:drm}
The diffusion process works in two steps: the forward process and the reverse process.

\vspace{+0.05cm}
\noindent \textbf{(1) Forward process:} The diffusion model corrupts the original data $\boldsymbol{x}_0$ by repeatedly adding $T$ Gaussian noises.\label{sec:diffusion_forward}
% \begin{align}
% & q(\boldsymbol{x}_t|\boldsymbol{x}_{t-1}) = \mathcal{N}(\boldsymbol{x}_t|\sqrt{1-\beta_t}\boldsymbol{x}_{t-1}, \beta_t\textbf{I}), \nonumber \\
% & q(\boldsymbol{x}_t|\boldsymbol{x}_0)=\mathcal{N}(\boldsymbol{x}_t|\sqrt{\bar{\alpha}_t}\boldsymbol{x}_0, (1-\bar{\alpha}_t)\textbf{I}),
% \end{align}
\begin{equation}
    q(\boldsymbol{x}_t|\boldsymbol{x}_{t-1}) = \mathcal{N}(\boldsymbol{x}_t|\sqrt{1-\beta_t}\boldsymbol{x}_{t-1}, \beta_t\textbf{I}),
\label{eq:forward}
\end{equation}
where $\beta_t \in [0,1]$ is a hyperparameter.
% $\alpha_t = 1-\beta_t$ and $\bar{\alpha}_t = \prod_{s=1}^t\alpha_s$.
% We can sample $\boldsymbol{x}_t$ at arbitrary step $t$ with a closed form by using the reparameterization trick \cite{vae}.

\vspace{+0.05cm}
\noindent \textbf{(2) Reverse process:} 
The ground-truth distribution of the reverse process requires the original data $\boldsymbol{x}_0$.
\begin{align}
q(\boldsymbol{x}_{t-1}|\boldsymbol{x}_t,\boldsymbol{x}_0) & = \mathcal{N}(\boldsymbol{x}_{t-1} | \boldsymbol{\mu}(\boldsymbol{x}_t, \boldsymbol{x}_0, t), \sigma^2(t)\textbf{I}), \text{ where} \nonumber\\
\boldsymbol{\mu}(\boldsymbol{x}_t, \boldsymbol{x}_0, t) & = \frac{\sqrt{\bar{\alpha}_{t-1}}\beta_t}{1-\bar{\alpha}_t}\boldsymbol{x}_0 + \frac{\sqrt{\alpha_t}(1-\bar{\alpha}_{t-1})}{1-\bar{\alpha}_t}\boldsymbol{x}_t, \nonumber\\
\sigma^2(t) & = \frac{(1-\alpha_t)(1-\bar{\alpha}_{t-1})}{1-\bar{\alpha}_{t}}.
\label{eq:ddpmrev_gt}
\end{align}
$\alpha_t = 1-\beta_t$ and $\bar{\alpha}_t = \prod_{s=1}^t\alpha_s$.
The diffusion model aims to estimate the above reverse process without the original data $\boldsymbol{x}_0$:
\begin{equation}
p_\theta(\boldsymbol{x}_{t-1}|\boldsymbol{x}_t) = \mathcal{N}(\boldsymbol{x}_{t-1} | \boldsymbol{\mu}_\theta(\boldsymbol{x}_t, t), \boldsymbol{\Sigma}_\theta(\boldsymbol{x}_t, t)), \nonumber
\label{eq:ddpmrev}
\end{equation}
where $\boldsymbol{\mu}_\theta$ and $\boldsymbol{\Sigma}_\theta$ are the mean and covariance predicted by neural networks parameterized by $\theta$.
% The network is tailored to specific tasks, like U-Net for image generation \cite{DDPM} and Transformer for text generation \cite{Diffusion_LM}.
% , to predict $\boldsymbol{x}_0$ from $\boldsymbol{x}_t$.

\begin{figure*}[t]
  \centering
  \includegraphics[width=\linewidth]{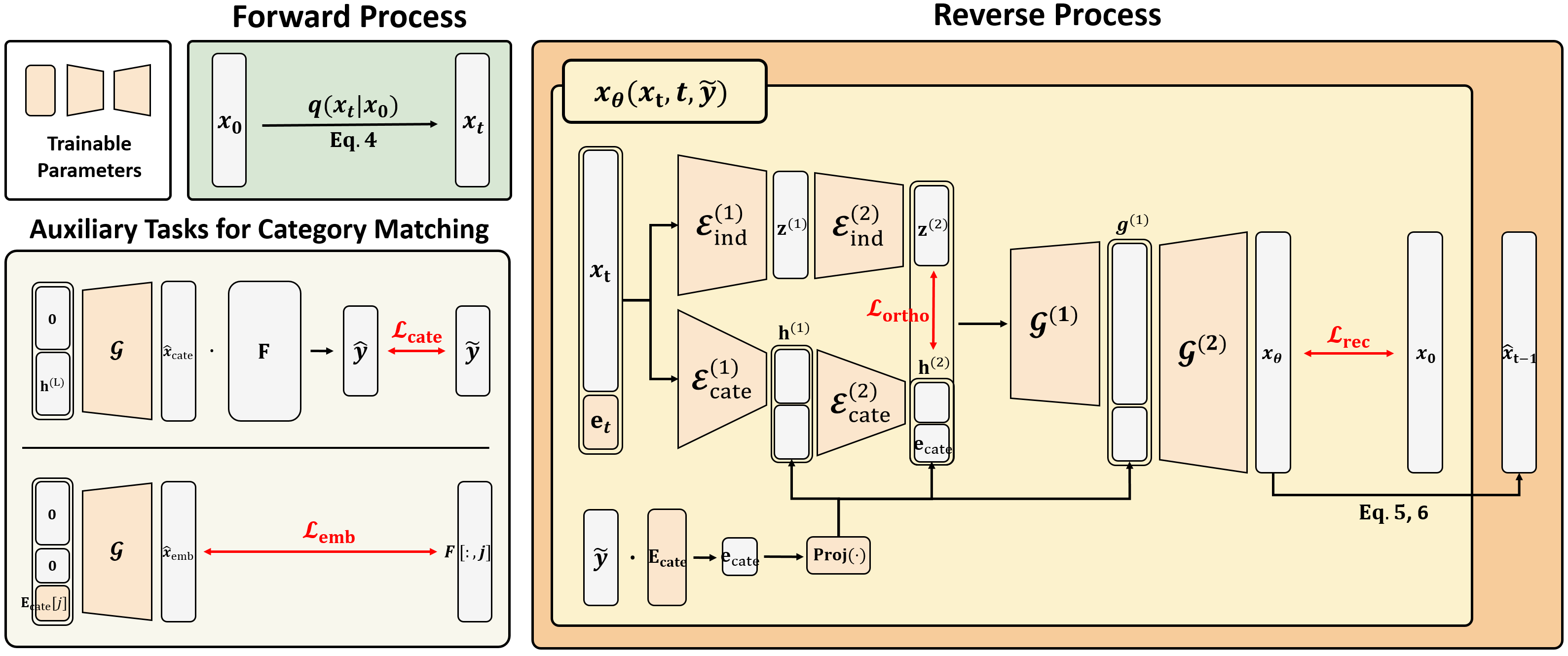}
  \vspace{+0.01cm}
  \caption{
  The overall framework of D3Rec.
  In the forward process, the user interactions are corrupted, thereby diminishing category preferences lurking in them.
  In the reverse process, the model generates the interactions guided by the targeted category preferences.
  In Section \ref{model:ae}, we take a closer look at $\mathcal{L}_\text{ortho}$ and $\mathcal{L}_\text{recon}$.
  In Section \ref{sec:refine_cate}, we devise two auxiliary tasks with $\mathcal{L}_\text{cate}$ and $\mathcal{L}_\text{emb}$ to ensure that the generated recommendations align with the targeted category preferences.
  }
  \label{fig:all}
  \vspace{-0.4cm}
\end{figure*}

\vspace{+0.05cm}
\noindent \textbf{(3) Optimization:} \label{sec:dif_opt}
The model is optimized by maximizing the Evidence Lower BOund (ELBO) of the log-likelihood $p(\boldsymbol{x}_0)$ \cite{DDPM_under}.
\begin{align}
\text{log} p(\boldsymbol{x}_0) 
&\ge \underbrace{\mathbb{E}_{q(\boldsymbol{x}_1|\boldsymbol{x}_0)}[ \text{log} p_\theta (\boldsymbol{x}_0|\boldsymbol{x}_1)]}_\text{reconstruction term} - \underbrace{KL(q(\boldsymbol{x}_T|\boldsymbol{x}_0)||p(\boldsymbol{x}_T))}_\text{prior matching term (const.)} \nonumber \\
& -\sum_{t=2}^T \underbrace{\mathbb{E}_{q(\boldsymbol{x}_t|\boldsymbol{x}_0)}[KL(q(\boldsymbol{x}_{t-1}|\boldsymbol{x}_t, \boldsymbol{x}_0)||p_\theta(\boldsymbol{x}_{t-1}|\boldsymbol{x}_t))]}_\text{denoising matching term}. 
\label{eq:elbo}
\end{align}
\noindent The ELBO is approximated using a Monte Carlo estimate as done for vanilla VAEs \cite{vae}.
% The prior matching term has no trainable parameters, thus it is constant in the optimization process and could be ignored.
% The denoising matching term makes $p_\theta(\boldsymbol{x}_{t-1}|\boldsymbol{x}_t)$ close to ground-truth transition step $q(\boldsymbol{x}_{t-1}|\boldsymbol{x}_t, \boldsymbol{x}_0)$.

\vspace{+0.05cm}
\noindent \textbf{(4) Inference:}
After training $\boldsymbol{\mu}_\theta$ and $\boldsymbol{\Sigma}_\theta$, $\boldsymbol{x}_T$ is denoised step-by-step $\boldsymbol{x}_T \rightarrow \boldsymbol{x}_{T-1} \rightarrow \cdots \rightarrow \boldsymbol{x}_0$ by utilizing $p_\theta(\boldsymbol{x}_{t-1}|\boldsymbol{x}_t)$.
Some studies control the generation process by adding conditions to obtain the desired $\boldsymbol{x}_0$ \cite{stable, dreamrec}.

\section{D3Rec}
We propose D3Rec (\underline{D}isentangled \underline{D}iffusion model for \underline{D}iversified \underline{Rec}ommendation), a framework for controlling recommendation lists based on arbitrary targeted category preferences.

\subsection{Overall Framework} \label{sec:overal_frame}
% First, we briefly present the overall framework (Section \ref{sec:overal_frame}), then describe the details of D3Rec (Figure \ref{fig:all}).
% To predict user interactions while reflecting diversity with the category preference, we disentangle the user representation into diversity-aware and diversity-independent features (Section \ref{sec:two_tower}) and refine the category preference representation (Section \ref{sec:refine_cate}).
% Afterward, to optimize the parameter of D3Rec, we define a loss function for training (Section \ref{sec:d3rec_opt}).
% Finally, we generate user interactions guided by the controlled category preferences, utilizing the classifier-free diffusion framework \cite{classifierfree} to rapidly adapt to the desired category preferences distribution (Section \ref{sec:d3rec_inference}).
D3Rec has two processes as shown in Figure \ref{fig:all}:
(1) D3Rec eliminates the outdated category preference information embedded in user interactions by gradually corrupting the interactions in the forward process.
(2) D3Rec generates users' future interactions from the corrupted interactions, considering the targeted category preferences in the reverse process.

\subsubsection{\textbf{Forward process to eliminate the outdated category preferences.}}
We set the initial state as $\boldsymbol{x}_0 = \boldsymbol{x}$ for a user $u$.\footnote{For simplicity in notation, we will omit the user index $u$ from $\boldsymbol{x}^u$.}
During the forward process, the historical interactions are corrupted by repeatedly adding Gaussian noises: 
% As discussed in Section \ref{sec:diffusion_forward}, $\boldsymbol{x}_t$ at step $t$ is formulated as:

\begin{align}
q(\boldsymbol{x}_t|\boldsymbol{x}_0)=\mathcal{N}(\boldsymbol{x}_t;\sqrt{\bar{\alpha}_t}\boldsymbol{x}_0, (1-\bar{\alpha}_t)\textbf{I}).
\end{align}
% where $\alpha_t = 1-\beta_t$ and $\bar{\alpha}_t = \prod_{s=1}^t\alpha_s$.
We can sample $\boldsymbol{x}_t$ at arbitrary step $t \in [0,T]$ with a closed form by using the reparameterization trick \cite{vae} and Eq.\ref{eq:forward}.
We adopt a linear schedule for $1 - \bar{\alpha}_t$ and $T$ is a hyper-parameter selected to be less than 100.
%and choose not to corrupt users' interactions into a standard Gaussian distribution to preserve personalized information. Among various beta schedules, according to \cite{DiffRec}, w

\subsubsection{\textbf{Reverse process guided by targeted category preferences.}}
Since the reverse process of DiffRec \cite{DiffRec} (Eq.\ref{eq:ddpmrev}) cannot reflect personal category preferences, we propose guiding the denoising transition with the targeted category preference $\tilde{\boldsymbol{y}} \in \mathbb{R}^{|\mathcal{C}|}$:
\begin{equation}
    p_\theta(\boldsymbol{x}_{t-1}|\boldsymbol{x}_t, \tilde{\boldsymbol{y}}) = \mathcal{N}(\boldsymbol{x}_{t-1} | \boldsymbol{\mu}_\theta(\boldsymbol{x}_t, t, \tilde{\boldsymbol{y}}), \boldsymbol{\Sigma}_\theta(\boldsymbol{x}_t, t, \tilde{\boldsymbol{y}})). \nonumber
\end{equation}
Here, the targeted category preference $\tilde{\boldsymbol{y}}$ can be any arbitrary vector that sums up to 1.
During the training, $\tilde{\boldsymbol{y}}$ is set to the original category preference $\boldsymbol{y}$.
We set the covariance constant as $\boldsymbol{\Sigma}_\theta(\boldsymbol{x}_t, t, \tilde{\boldsymbol{y}}) = \sigma^2(t)\boldsymbol{I}$ in Eq \ref{eq:ddpmrev_gt} for stable training, following \cite{DDPM, DiffRec}.
% $\sigma^2(t)\boldsymbol{I}$ represents the covariance of the ground-truth distribution $q(\boldsymbol{x}_{t-1}|\boldsymbol{x}_t, \boldsymbol{x}_0)$.
We model $\boldsymbol{\mu}_\theta(\boldsymbol{x}_t, t, \tilde{\boldsymbol{y}})$ in a similar form of the mean in Eq \ref{eq:ddpmrev_gt}:
\begin{align}
\boldsymbol{\mu}_\theta(\boldsymbol{x}_t, t, \tilde{\boldsymbol{y}}) 
        = \frac{\sqrt{\bar{\alpha}_{t-1}}\beta_t}{1-\bar{\alpha}_t}\tilde{\boldsymbol{x}}_\theta(\boldsymbol{x}_t, t , \tilde{\boldsymbol{y}}) + \frac{\sqrt{\alpha_t}(1-\bar{\alpha}_{t-1})}{1-\bar{\alpha}_t}\boldsymbol{x}_t \label{eq:d3rec_rev},
\end{align}
where $\tilde{\boldsymbol{x}}_\theta (\boldsymbol{x}_t,t,\tilde{\boldsymbol{y}})$ is the model prediction for $\boldsymbol{x}_0$ and defined with the classifier-free diffusion guidance \cite{classifierfree}:
\begin{align}
\tilde{\boldsymbol{x}}_\theta(\boldsymbol{x}_t, t, \tilde{\boldsymbol{y}}) = (1+w) \cdot \boldsymbol{x}_\theta(\boldsymbol{x}_t, t, \tilde{\boldsymbol{y}}) - w \cdot \boldsymbol{x}_\theta(\boldsymbol{x}_t, t, \boldsymbol{0}^{|\mathcal{C}|}). \label{eq:pred_conditioned}
\end{align}
$w$ is a hyper-parameter that determines the balance between category-guided prediction $\boldsymbol{x}_\theta(\boldsymbol{x}_t, t, \tilde{\boldsymbol{y}})$ and unconditional prediction $ \boldsymbol{x}_\theta(\boldsymbol{x}_t, t, \boldsymbol{0}^{|\mathcal{C}|})$.
$\boldsymbol{0}^{|\mathcal{C}|}$ is a $|\mathcal{C}|$-dimensional zero vector to feed unconditional state to the model.
% To train the unconditional model, we randomly drop out the condition with a tunable probability.

\subsection{Proposed Architecture for Reverse Process} \label{model:ae}
% To compute $\tilde{\boldsymbol{x}}_\theta(\cdot)$ in Eq. \ref{eq:pred_conditioned}, we propose a model $\boldsymbol{x}_\theta$ depicted in Figure \ref{fig:all}.
$\boldsymbol{x}_\theta(\boldsymbol{x}_t, t, \tilde{\boldsymbol{y}})$ in Eq. \ref{eq:pred_conditioned} is the only trainable model in our D3Rec framework.
In the following section, we present our architecture for $\boldsymbol{x}_\theta(\boldsymbol{x}_t, t, \tilde{\boldsymbol{y}})$ to generate $\boldsymbol{x}_0$ with $(\boldsymbol{x}_t, t, \tilde{\boldsymbol{y}})$.

% \noindent \textbf{(1) Disentangled two-tower encoders:}
% One tower $\mathcal{E}_{\text{cate}}$ encodes the category preference according to $\tilde{\boldsymbol{y}}$ and the other tower $\mathcal{E}_{\text{ind}}$ extracts a latent vector that captures category-independent factors, such as item's quality, etc \cite{DCRS}.

% \noindent \textbf{(2) Category preference-guided decoder:}
% The decoder $\mathcal{G}$ reconstructs user interactions by taking $\tilde{\boldsymbol{y}}$ into consideration.
% Note that, we set $\tilde{\boldsymbol{y}} = \boldsymbol{y}$ during training since the category preferences distribution of denoised user interactions $\boldsymbol{x}_0$ aligns with $\tilde{\boldsymbol{y}}$.
% % $\tau = 1$ during training, i.e., $\tilde{\boldsymbol{y}} = \boldsymbol{y}$, 

\subsubsection{\textbf{Category preference representation}} \label{sec:cate_pref}
We first define a category embedding matrix $\boldsymbol{E}_{\text{cate}} \in \mathbb{R}^{|\mathcal{C}| \times d}$. %, where $d$ is the output dimension of the encoder $\mathcal{E}_{\text{cate}}$
Then, the category preference representation $\boldsymbol{e}_\text{cate} \in \mathbb{R}^{d}$ is computed as a multiplication between the category preference vector and the category embedding matrix: $\boldsymbol{e}_\text{cate} = \boldsymbol{E}_{\text{cate}}^\top \tilde{\boldsymbol{y}} $.
% \begin{align}
%     \boldsymbol{e}_\text{cate} = \tilde{\boldsymbol{y}} \cdot \boldsymbol{E}_{\text{cate}}.
% \end{align}

\subsubsection{\textbf{Disentangled two-tower encoders}} \label{sec:two_tower}
We devise two-tower encoders: a category preference encoder $\mathcal{E}_{\text{cate}}$ and a category-independent encoder $\mathcal{E}_{\text{ind}}$.
Both encoders are comprised of Multi-Layer Perceptron (MLP) with $L$ layers and the input dimension is identical.
The output dimension of $\mathcal{E}_{\text{cate}}$ is $d$ which is half of $\mathcal{E}_{\text{ind}}$. 
The remaining output dimensions in $\mathcal{E}_{\text{cate}}$ are allocated for the category preference representation $\boldsymbol{e}_{\text{cate}} \in \mathbb{R}^d$.

\vspace{+0.1cm}
\noindent \textbf{Category preference encoder $\mathcal{E}_{\text{cate}}$.}
We provide $\boldsymbol{e}_\text{cate}$ to each layer of $\mathcal{E}_{\text{cate}}$ to capture category-aware features, as follows:
\begin{align}
    &\boldsymbol{h}^{(l)} = \text{CONCAT}\big(\mathcal{E}_{\text{cate}}^{(l)} \big( \boldsymbol{h}^{(l-1)} \big); \text{Proj}^{(l)}(\boldsymbol{e}_\text{cate}) \big), \nonumber \\
    &\boldsymbol{h}^{(L)} = \text{CONCAT}\big(\mathcal{E}_{\text{cate}}^{(L)} \big( \boldsymbol{h}^{(L-1)} \big); \boldsymbol{e}_\text{cate} \big) \in \mathbb{R}^{2d}.
\label{eq:cate_enc}
\end{align}
$\text{Proj}^{(l)}$ refers to a 1-layer MLP designed to project $\boldsymbol{e}_\text{cate}$ into the manifold space of each layer.
The input is $\boldsymbol{h}^{(0)} = \text{CONCAT}(\boldsymbol{x}_t;\boldsymbol{e}_t)$, where $\boldsymbol{e}_t$ is the positional embedding of step $t$ \cite{transformer}.

%, and $\text{sg}(\cdot)$ stands for the stop-gradient operator \cite{vqvae}, which is defined as having an identity function during forward computation but having zero partial derivatives.
% Let's define $d^{(l)}$ as the output dimension of layer $l$ in the encoder $\mathcal{E}_\delta$, described in Section \ref{sec:d3rec_idp}.
% Then, $\boldsymbol{h}^{(l)} \in \mathbb{R}^{d^{(l)}}$ and $\boldsymbol{h}_q^{(l)}, \boldsymbol{h}^{(l)}_p \in \mathbb{R}^{d^{(l)} / 2}$. 
% The term $\boldsymbol{h}_q$ embodies arbitrary knowledge linked to diversity, which may be influenced by $\boldsymbol{h}_p$, but does not precisely align with it, such as tolerance for boredom, etc.
% $\boldsymbol{e}_\text{cate}$ represents the category preferences and 
% To maintain category preference information and training stability, we enforce $\boldsymbol{h}_p^{(l)}$ as a non-updated constant by applying the stop-gradient operator $\text{sg}(\cdot)$.
% Further details about the training of $\text{e}_{c_j}$ and $\text{Proj}(\cdot)$ are discussed in Section \ref{sec:refine_cate}.

\vspace{+0.1cm}
\noindent \textbf{Category-independent encoder $\mathcal{E}_{\text{ind}}$.} \label{sec:d3rec_idp}
We do not utilize the category preference representation for $\mathcal{E}_{\text{ind}}$:
\begin{equation}
\boldsymbol{z}^{(L)} = \mathcal{E}_{\text{ind}}(\boldsymbol{z}^{(0)}) \ \in \mathbb{R}^{2d}, \nonumber
\end{equation}
where the input is $\boldsymbol{z}^{(0)} = \boldsymbol{h}^{(0)} = \text{CONCAT}(\boldsymbol{x}_t;\boldsymbol{e}_t)$.
% The encoder $\mathcal{E}_{\text{ind}}$ is aimed to extract a diversity-independent feature $\boldsymbol{z}$ from $\boldsymbol{x}_t$. 
To extract category-independent features, we employ orthogonal disentanglement \cite{ortho_ae}: %.After obtaining $\boldsymbol{z}^{(L)}$ and $\boldsymbol{h}^{(L)}$, we impose orthogonality by minimizing the cosine similarity between them, expressed as
\begin{equation}
    \mathcal{L}_\text{ortho} = \text{Cosine}\big( \boldsymbol{z}^{(L)}, \boldsymbol{h}^{(L)}\big), \label{eq:loss_ortho}
\end{equation}
where $\text{Cosine}(\cdot,\cdot)$ is the cosine similarity.
%, and $\boldsymbol{z}^{(l)} \in \mathbb{R}^{d^{(l)}}$ for $l \in [1, L]$.
Disentangling the category representation $\boldsymbol{h}^{(L)}$ is essential for diversity control, as it allows for the manipulation of category preferences while preserving the user's category-independent tastes.

\subsubsection{\textbf{Category preference-guided decoder}}
The input of the decoder $\mathcal{G}$ is the concatenation of the outputs from the encoders $\mathcal{E}_{\text{cate}}$ and $\mathcal{E}_{\text{ind}}$.
In the context of predicting denoised interactions based on $\tilde{\boldsymbol{y}}$, similar to the approach of $\mathcal{E}_{\text{cate}}$, we provide the representation $\boldsymbol{e}_\text{cate}$ to each intermediate layer of $\mathcal{G}$:
% \begin{alignat}{2}
%     &\boldsymbol{g}^{(l)} = \begin{cases}
%             \boldsymbol{g}^{(l)}_q \\
%             \text{CONCAT}(\boldsymbol{g}^{(l)}_q, \boldsymbol{g}^{(l)}_p)
%         \end{cases}
%         &&\def\arraystretch{1.2}\begin{array}{@{}l}
%         ,\text{if } l = L \\
%         ,\text{otherwise} \\
%         \end{array}
%          \\
%     \nonumber \\
%     &
%     \boldsymbol{g}^{(l)}_q = \mathcal{G}^{(l)}(\boldsymbol{g}^{(l-1)})
%     \ \text{ and } \
%     \boldsymbol{g}^{(l)}_p = \boldsymbol{h}&&^{(L-l)}_p
%     \nonumber
% \end{alignat}
\begin{equation}
    \boldsymbol{g}^{(l)} = \text{CONCAT}(\mathcal{G}^{(l)}(\boldsymbol{g}^{(l-1)}); \text{Proj}^{(L-l)}(\boldsymbol{e}_\text{cate})) \nonumber
\end{equation}
for $l \in [1, L]$ and $\boldsymbol{g}^{(0)} =  \text{CONCAT}(\boldsymbol{z}^{(L)};\boldsymbol{h}^{(L)})$.
% Each representation's dimension is $\boldsymbol{g}^{(l)}_q \in \mathbb{R}^{3d^{(L-l)}/2}$, $\boldsymbol{g}^{(l)}_p \in  \mathbb{R}^{d^{(L-l)}/2}$, and $\boldsymbol{g}^{(L)} \in \mathbb{R}^{|\mathcal{I}|}$.
Note that we utilize $\text{Proj}^{(L-l)}(\boldsymbol{e}_\text{cate})$ for $\mathcal{G}^{(l)}$ since the encoder-decoder architecture is similar to U-Net \cite{Unet}.
The output $\boldsymbol{g}^{(L)} \in \mathbb{R}^{|\mathcal{I}|}$ is the generated user interactions, represented by $\boldsymbol{x}_\theta(\boldsymbol{x}_t, t, \tilde{\boldsymbol{y}}) = \boldsymbol{g}^{(L)}$.

\subsubsection{\textbf{Maximization of ELBO conditioned $\boldsymbol{y}$}}
The primary training objective is to maximize the ELBO of observed user interactions $\boldsymbol{x}_0$ while taking into account the original category preference $\boldsymbol{y}$.
% \begin{align} \label{eq:elbo_d3rec}
%     \text{log}p(\boldsymbol{x}_0) &\ge 
%     \underbrace{\mathbb{E}_{q(\boldsymbol{x}_1|\boldsymbol{x}_0)}[ \text{log} p_\theta (\boldsymbol{x}_0|\boldsymbol{x}_1, \boldsymbol{y})]}_\text{reconstruction term} \\
%     & -\sum_{t=2}^T \underbrace{\mathbb{E}_{q(\boldsymbol{x}_t|\boldsymbol{x}_0)}[KL(q(\boldsymbol{x}_{t-1}|\boldsymbol{x}_t, \boldsymbol{x}_0)||p_\theta(\boldsymbol{x}_{t-1}|\boldsymbol{x}_t, \boldsymbol{y}))]}_\text{denoising matching term}. \nonumber
% \end{align}
% \noindent Note that the prior matching term in Eq.\ref{eq:elbo} is excluded because it is a constant.
% Afterward, as discussed in \cite{multvae}, we estimate the negative of the reconstruction term.
Using Gaussian transition assumption and Bayes' theorem, the ELBO in Eq.\ref{eq:elbo} can be simplified into following reconstruction loss \cite{DDPM, DDPM_under}:
% \begin{alignat}{2}
%     &\mathcal{L}_\text{recon} = \begin{cases}
%             \ \ \ \ ||\boldsymbol{x}_\theta(\boldsymbol{x}_t, t, \boldsymbol{y}) - \boldsymbol{x}_0||_2^2 \\
%             \ \beta ||\boldsymbol{x}_\theta(\boldsymbol{x}_t, t, \boldsymbol{y})-\boldsymbol{x}_0||^2_2
%         \end{cases}
%         &&\def\arraystretch{1.2}\begin{array}{@{}l}
%         ,\text{if } t = 1 \\
%         ,\text{otherwise}
%         \end{array}
%     \label{eq:loss_recon}
% \end{alignat}
% \noindent where $\beta = \frac{1}{2}(\text{SNR}(t-1) - \text{SNR}(t))$, $\text{SNR}(t) = \bar{\alpha}_t / (1 - \bar{\alpha}_t)$. 
% \begin{equation}
%     \mathcal{L}_\text{recon} = \frac{\delta}{|\mathcal{I}|} \sum_{i \in \mathcal{I}} (\boldsymbol{x}_\theta(\boldsymbol{x}_t, t, \boldsymbol{y})_i - x_i)^2
% \label{eq:loss_recon}
% \end{equation}
\begin{equation}
    \mathcal{L}_\text{recon} = \delta \cdot ||\boldsymbol{x}_\theta(\boldsymbol{x}_t, t, \boldsymbol{y})-\boldsymbol{x}_0||^2_2,
\label{eq:loss_recon}
\end{equation}
where $\delta =1$ when $t=1$, and $\delta = \frac{1}{2}(\text{SNR}(t-1) - \text{SNR}(t))$ otherwise.
$\text{SNR}(t) = \bar{\alpha}_t / (1 - \bar{\alpha}_t)$ is the signal-noise ratio.

\subsection{Auxiliary Tasks for Category Matching}\label{sec:refine_cate}
We devise two auxiliary tasks to ensure that the generated recommendations align with the targeted category preferences.

% \noindent \textbf{1) Predicting the category preferences.}
% 1) Predicting the category preferences $\boldsymbol{y}$ using $\boldsymbol{e}_\text{cate}$ and $\text{Proj}^{(l)}$ alongside $\boldsymbol{h}_q^{(L)}$, and 2) Quantifying the association between each category and the items belonging to that particular category, using $\boldsymbol{e}_{c_j}$.
\subsubsection{\textbf{Predicting the category preferences}}
The category-aware encoder $\mathcal{E}_{\text{cate}}$ should capture the category preference $\boldsymbol{y}$ by itself.
To this end, we provide only $\boldsymbol{h}^{(L)}$ to the decoder by replacing $\boldsymbol{z}^{(L)}$ with $\boldsymbol{0}^{2d}$:
\begin{equation}
    \boldsymbol{g}_\text{cate}^{(0)} = \text{CONCAT}(\boldsymbol{0}^{2d};\boldsymbol{h}^{(L)}), \nonumber
\end{equation}
and obtain the output $\boldsymbol{g}_\text{cate}^{(L)} = \hat{\boldsymbol{x}}_\text{cate} \in \mathbb{R}^{|\mathcal{I}|}$.
Then, we estimate the category preference vector with the reconstructed user history.
\begin{align}
& \hat{\boldsymbol{y}} = \boldsymbol{F}^\top \hat{\boldsymbol{x}}_\text{cate} \ / \ ||\boldsymbol{F}^\top \hat{\boldsymbol{x}}_\text{cate}||_1, \nonumber\\
&\mathcal{L}_\text{cate} = || \hat{\boldsymbol{y}} - \tilde{\boldsymbol{y}} ||_2^2. \label{eq:loss_cate}
\end{align}
$\boldsymbol{F} \in \mathbb{R}^{|\mathcal{I}| \times |\mathcal{C}|}$ is the item-category matrix defined in Section \ref{sec:notation}.
% $\hat{\boldsymbol{x}}_\text{cate}[i] \in \mathbb{R}$ is the predicted interaction for item $i \in \mathcal{I}$ and

\subsubsection{\textbf{Association between category embeddings and items}}
For each category $j \in \mathcal{C}$, its embedding $\boldsymbol{E}_{\text{cate}}[j] \in \mathbb{R}^{d}$ should be associated with items belonging to that particular category.
To this end, we provide only $\boldsymbol{E}_{\text{cate}}[j]$ to the decoder:
\begin{equation}
    \boldsymbol{g}_\text{emb}^{(0)} = \text{CONCAT}(\boldsymbol{0}^{2d};\boldsymbol{0}^{d};\boldsymbol{E}_{\text{cate}}[j]), \nonumber
\end{equation}
by replacing $\boldsymbol{z}^{(L)}$ with $\boldsymbol{0}^{2d}$ and $\boldsymbol{h}^{(L)}$ in Eq.\ref{eq:cate_enc} with $\text{CONCAT}(\boldsymbol{0}^{d}; \boldsymbol{E}_{\text{cate}}[j])$.
Then, the reconstructed user history $\boldsymbol{g}_\text{emb}^{(L)} = \hat{\boldsymbol{x}}_\text{emb} \in \mathbb{R}^{|\mathcal{I}|}$ should indicate the items belonging to the category $j$:
\begin{align}
\mathcal{L}_\text{emb} = \sum_{j \in \mathcal{C}}||\hat{\boldsymbol{x}}_{\text{emb}} - F[:, j] ||_2^2, \label{eq:loss_emb}
\end{align}
where $\boldsymbol{F}[:, j] \in \mathbb{R}^{|\mathcal{I}|}$ denotes the $j$-th column of the item-category matrix $\boldsymbol{F} \in \mathbb{R}^{|\mathcal{I}| \times |\mathcal{C}|}$.
%and $\hat{\boldsymbol{x}}_{emb} = \boldsymbol{g}_{c_j}^{(L)}$.
% Note that we use $\boldsymbol{e}_{c_j}$ instead of $\boldsymbol{e}_\text{cate}$ in this stage.

\subsection{Model Training} \label{sec:d3rec_opt}
% $\mathcal{L}_1 \approx ||\boldsymbol{x}_\theta(\boldsymbol{x}_1, 1, \boldsymbol{y}) - \boldsymbol{x}_0||_2^2$

% as $\mathcal{L}_t=\frac{1}{2}(\text{SNR}(t-1) - \text{SNR}(t)) \big[||\boldsymbol{x}_\theta(\boldsymbol{x}_t, t, \boldsymbol{y})-\boldsymbol{x}_0||^2_2 \big]$, where $\text{SNR}(t) = \bar{\alpha}_t / (1 - \bar{\alpha}_t)$.
% Therefore, for maximizing the Eq.\ref{eq:elbo_d3rec}, we optimizing $\boldsymbol{x}_\theta(\boldsymbol{x}_t, t, \boldsymbol{y})$ by minimizing the reconstruction loss:

% \begin{equation}
%     \mathcal{L}_\text{recon} = \sum_{t=1}^T \mathcal{L}_t \label{eq:loss_recon}
% \end{equation}

\subsubsection{\textbf{Re-weight Strategy}}
\label{sec:reweight}
The uneven distribution of user interactions across categories introduces an imbalance among categories in the training data, leading to popular categories dominating the training gradient.
% This phenomenon diminishes the diversity of the recommended list.
% Inspired by the effective resolution \cite{focal} of class imbalance issues, 
We propose a re-weight strategy that assigns more weight to hard examples and less weight to easy examples, inspired by Focal loss \cite{focal}.
Given $\boldsymbol{y}$, we devise two category weight vectors:
\begin{align}
  & \boldsymbol{y}^\text{pos} = \gamma_\text{min} + (\gamma_\text{max} - \gamma_\text{min}) \cdot \frac{(1-\boldsymbol{y}) - \text{min}(1-\boldsymbol{y})}{\text{max}(1-\boldsymbol{y}) - \text{min}(1-\boldsymbol{y})},\nonumber \\
  & \boldsymbol{y}^\text{neg} = \gamma_\text{min} + (\gamma_\text{max} - \gamma_\text{min}) \cdot \frac{\boldsymbol{y} - \text{min}(\boldsymbol{y})}{\text{max}(\boldsymbol{y}) - \text{min}(\boldsymbol{y})},\nonumber
\end{align}
% \noindent However, directly using $\boldsymbol{y}^\prime$ as weights can lead to certain issues.
% Assigning excessively large or small weights can reduce recommendation accuracy and hinder model convergence \cite{ddrn}. 
% We resolve this by applying linear normalization to ensure that the weights fall within an appropriate range $[\gamma_\text{min}, \gamma_\text{max}]$:
\noindent where $\gamma_\text{min} < \gamma_\text{max}$ control the upper and lower bounds of the weight.
We assign a weight $\rho_i \in \mathbb{R}$ for each item $i \in \mathcal{I}$ based on its category distribution $\boldsymbol{F}[i] \in \mathbb{R}^{|\mathcal{C}|}$ and interaction $\boldsymbol{x}_0[i] \in \{0,1\}$:
\begin{align}
&\rho_i = \boldsymbol{F}[i]^\top \boldsymbol{y}^\text{pos} \ \ \text{  if } \ \boldsymbol{x}_0[i]=1, \nonumber \\
&\rho_i = \boldsymbol{F}[i]^\top \boldsymbol{y}^\text{neg} \ \ \text{  if } \ \boldsymbol{x}_0[i]=0.
\label{eq:rhoo}
\end{align}
For positive samples, we assign higher weights to the loss of items belonging to minor categories, and lower weights to those belonging to major categories.
Conversely, for negative samples, we assign higher weights to the loss of items belonging to major categories, and lower weights to those belonging to minor categories.
Then, $\mathcal{L}_\text{recon}$ is re-weighted as follows:
\begin{align} 
    \text{re-weight}(\mathcal{L}_\text{recon}) = \delta \cdot ||\boldsymbol{\rho}^{0.5} \cdot (\boldsymbol{x}_\theta(\boldsymbol{x}_t, t, \boldsymbol{y})-\boldsymbol{x}_0)||^2_2,\label{eq:re_weight}
\end{align}
where $\boldsymbol{\rho} \in \mathbb{R}^{|\mathcal{I}|}$ is a weight vector with $\boldsymbol{\rho}[i] = \rho_i$ defined in Eq.\ref{eq:rhoo}.

\subsubsection{\textbf{Optimization objective}}
The conclusive optimization objective of D3Rec aggregates all loss functions described above:
% To achieve these tasks, according to Eq. \ref{eq:loss_recon}, Eq. \ref{eq:loss_ortho}, Eq. \ref{eq:loss_cate} and Eq. \ref{eq:loss_emb}, we optimize the model parameter $\theta$ by minimizing following objective:
\begin{equation}
    \mathbb{E}_{u \in \mathcal{U}, t \in [T]}\big[\text{re-weight}(\mathcal{L}_\text{recon}) + \mathcal{L}_\text{cate} + \lambda(\mathcal{L}_\text{ortho} + \mathcal{L}_\text{emb})\big],
\end{equation}
where $\lambda$ is a hyper-parameter that controls the effects of the each loss functions.

\subsection{Inference via Targeted Category Preference} \label{sec:d3rec_inference}
In the inference phase, we can utilize arbitrary targeted category preference $\tilde{\boldsymbol{y}}$ for $\boldsymbol{x}_{\theta}(\boldsymbol{x}_t, t, \tilde{\boldsymbol{y}})$.
If the original category preference $\boldsymbol{y}$ is utilized, D3Rec would output recommendations aligned with the original user category preference.
On the other hand, to control diversity, the category preference can be adjusted to the business strategy.
Broadly, the usages are divided into two directions.

\vspace{+0.1cm}
\noindent \textbf{Modifying the category preferences through temperature $\tau$.}
We regulate diversity while accommodating the user's general preference by deploying $\tilde{\boldsymbol{y}} = \text{Softmax}(\text{log}(\boldsymbol{y}) / \tau)$.\label{eq:control}
If $\tau < 1$, $\tilde{\boldsymbol{y}}$ skews towards the historically major categories, consequently reducing diversity in the recommended list.
On the other hand, when $\tau > 1$, $\tilde{\boldsymbol{y}}$ becomes smoother, thereby enhancing diversity within the list.

\vspace{+0.1cm}
\noindent \textbf{Manipulating the preferences into specific categories.}
The system or users can set arbitrary targeted preferences $\tilde{\boldsymbol{y}}$ that may deviate from the original preferences $\boldsymbol{y}$. This approach allows the recommendation system to adapt readily to the users' capricious moods (e.g., a user who typically prefers action movies might request romantic movies when with a partner).

\vspace{+0.1cm}
\noindent As outlined in \cite{multvae, DiffRec}, we disregard the covariance at inference.
Thus, we conduct deterministic inference i.e., $\boldsymbol{x}_{t-1} = \boldsymbol{\mu}_\theta(\boldsymbol{x}_t, t, \tilde{\boldsymbol{y}})$ in Eq. \ref{eq:d3rec_rev}.
Additionally, we corrupt $\boldsymbol{x}_0$ for $T^\prime$ steps and then denoise it for $T$ steps, where $T^\prime < T$.
This approach reflects the inherent noise in user interactions, such as false positives and false negatives, while also retaining personalized information.

% \section{Proposed model architecture} \label{model:ae}
% \input{5archi}

\section{Experiments}
% We experiment with D3Rec to validate the following points: 
% 1) Considering category information is beneficial for alleviating the accuracy-diversity dilemma.
% 2) Removing outdated category preference data from past interactions is crucial for controlling the recommendation system based on targeted category preferences.
We validate the superiority of D3Rec over state-of-the-art methods for diverse recommendations using real-world datasets.
We have released the anonymized source code.\footnote{\url{https://anonymous.4open.science/r/D3Rec-3EF7}}

\subsection{Experimental Setup}
We present detailed experimental setup in Appendix \ref{sec:detailexp}.
\begin{table}[t]
 \renewcommand{\arraystretch}{0.6}
  \caption{Data statistics.}
  \vspace{+0.15cm}
  \begin{tabular}{cccccc}
    \toprule
    Dataset & \#Users & \#Items & \#Interactions & \#Categories \\
    \midrule
    ML-1M &5,148 & 2,380 & 553,277 & 18 \\
    Steam Game & 31,572 & 6,070 & 1,351,659 & 12 \\
    Anime 2023 & 143,907 & 7,947 & 12,803,010 & 21 \\
    \bottomrule
  \end{tabular}
  \label{table:statistics}
  \vspace{-0.2cm}
\end{table}

\vspace{+0.1cm}
\noindent \textbf{Datasets.}
We adopt three real-world datasets in different domains, including \textbf{ML-1M}\footnote{https://grouplens.org/datasets/movielens/}, \textbf{Steam Game} \cite{steam}, and \textbf{Anime 2023}\footnote{https://www.kaggle.com/datasets/dbdmobile/myanimelist-dataset/data}.
For explicit datasets (ML-1M and Anime 2023), we convert ratings higher than the middle score to 1 and 0 otherwise.
For all datasets, we consider an item's genres as the categories.
We adopt 20-core settings for users, items, and categories across all datasets to ensure data quality \cite{DCRS}.
After that, we split each user’s interactions into a training set (60\%), a validation set (20\%), and a test set (20\%).
The statistics of the three datasets after the pre-processing are presented in Table \ref{table:statistics}.

% Please add the following required packages to your document preamble:
% \usepackage{multirow}
% \usepackage[normalem]{ulem}
% \useunder{\uline}{\ul}{}
\begin{table*}[t]
 \caption{
 The overall performance comparison between D3Rec and compared methods across three datasets. The best results are highlighted in bold and the second-best are underlined. \textit{Improv} (\%) denotes the relative performance improvement of D3Rec over the best result among the compared methods.  
 }
 \vspace{+0.1cm}
 \resizebox{\textwidth}{!}{
  \begin{tabular}{cl | cccc | cccc | cccc}
  \toprule
  \multirow{2}{*}{K} & \multicolumn{1}{c|}{\multirow{2}{*}{Models}} & \multicolumn{4}{c|}{ML-1M} & \multicolumn{4}{c|}{Steam Game} & \multicolumn{4}{c}{Anime 2023} \\
  & \multicolumn{1}{c|}{} & \multicolumn{1}{c}{Recall} & \multicolumn{1}{c}{NDCG} & \multicolumn{1}{c}{Entropy} & \multicolumn{1}{c|}{Coverage} & \multicolumn{1}{c}{Recall} & \multicolumn{1}{c}{NDCG} & \multicolumn{1}{c}{Entropy} & \multicolumn{1}{c|}{Coverage} & \multicolumn{1}{c}{Recall} & \multicolumn{1}{c}{NDCG} & \multicolumn{1}{c}{Entropy} & \multicolumn{1}{c}{Coverage} \\
\midrule
\midrule
  \multirow{12}{*}{10} & MultVAE & 0.0522 & 0.0884 & 0.6142 & 0.4867 & 0.0452 & 0.0523 & 0.6363 & 0.5929 & 0.0175 & 0.0230 & 0.6859 & 0.4966 \\
  & DiffRec & \underline{0.0547} & \underline{0.0989} & 0.6578 & 0.5300 & 0.0446 & 0.0511 & 0.6618 & \underline{0.6397} & \underline{0.0194} & \underline{0.0272} & 0.6934 & 0.5040 \\
  \cmidrule{2-14}
  & MacridVAE & 0.0459 & 0.0822 & 0.6450 & 0.5066 & 0.0413 & 0.0491 & 0.6556 & 0.6213 & 0.0184 & 0.0242 & 0.6973 & 0.5035 \\
  & DCRS & 0.0539 & 0.0931 & 0.6427 & 0.5116 & \underline{0.0465} & \underline{0.0552} & 0.6516 & 0.6027 & 0.0170 & 0.0252 & 0.6840 & 0.4969 \\
  \cmidrule{2-14}
  & COR & 0.0516 & 0.0929 & 0.6281 & 0.4993 & 0.0445 & 0.0546 & 0.6101 & 0.5728 & 0.0164 & 0.0242 & 0.6826 & 0.4977 \\
  & Dual Process & 0.0529 & 0.0921 & 0.6177 & 0.4880 & 0.0456 & 0.0551 & 0.6460 & 0.6039 & 0.0174 & 0.0239 & 0.6929 & 0.5065 \\
  & MMR & 0.0527 & 0.0891 & 0.6401 & 0.5188 & 0.0452 & 0.0521 & 0.6368 & 0.5935 & 0.0172 & 0.0221 & 0.6965 & 0.5104 \\
  & PMF & 0.0504 & 0.0764 & 0.6325 & 0.5074 & 0.0417 & 0.0409 & 0.6396 & 0.5993 & 0.0157 & 0.0185 & 0.6808 & 0.4927 \\
  & DPP & 0.0529 & 0.0893 & 0.6562 & 0.5408 & 0.0448 & 0.0517 & 0.6377 & 0.5951 & 0.0164 & 0.0212 & 0.7002 & 0.5172 \\
  & CATE & 0.0526 & 0.0910 & \underline{0.6743} & \underline{0.5426} & 0.0455 & 0.0522 & \underline{0.6779} & 0.6146 & 0.0169 & 0.0221 & \underline{0.7140} & \underline{0.5197} \\
  \cmidrule{2-14}
  & D3Rec & \textbf{0.0567} & \textbf{0.1006} & \textbf{0.7259} & \textbf{0.5919} & \textbf{0.0534} & \textbf{0.0659} & \textbf{0.7329} & \textbf{0.7188} & \textbf{0.0232} & \textbf{0.0322} & \textbf{0.7364} & \textbf{0.5553} \\
  & \textit{Improv} (\%) & +3.55 & +1.71 & +7.65 & +9.10 & +14.98 & +19.38 & +8.12 & +12.36 & +19.70 & +18.22 & +3.13 & +6.84 \\
\midrule
  \multirow{12}{*}{20} & MultVAE & 0.0921 & 0.0977 & 0.6763 & 0.6383 & 0.0724 & 0.0617 & 0.6719 & 0.6936 & 0.0280 & 0.0259 & 0.7341 & 0.5987 \\
  & DiffRec & 0.0949 & \underline{0.1062} & 0.7099 & 0.6703 & 0.0710 & 0.0605 & 0.6932 & 0.7389 & \underline{0.0313} & \underline{0.0302} & 0.7360 & 0.6005 \\
  \cmidrule{2-14}
  & MacridVAE & 0.0832 & 0.0893 & 0.7016 & 0.6577 & 0.0680 & 0.0576 & 0.6902 & 0.7333 & 0.0301 & 0.0276 & 0.7374 & 0.5922 \\
  & DCRS & \underline{0.0956} & 0.1024 & 0.6998 & 0.6537 & \underline{0.0734} & \underline{0.0640} & 0.6823 & 0.6944 & 0.0280 & 0.0278 & 0.7319 & 0.5933 \\
  \cmidrule{2-14}
  & COR & 0.0901 & 0.1002 & 0.6831 & 0.6414 & 0.0707 & 0.0623 & 0.6415 & 0.6646 & 0.0266 & 0.0265 & 0.7275 & 0.5931 \\
  & Dual Process & 0.0936 & 0.1013 & 0.6812 & 0.6371 & 0.0710 & 0.0629 & 0.6791 & 0.6981 & 0.0279 & 0.0266 & 0.7369 & 0.6024 \\
  & MMR & 0.0934 & 0.0987 & 0.6921 & 0.6621 & 0.0724 & 0.0616 & 0.6724 & 0.6942 & 0.0279 & 0.0254 & 0.7402 & 0.6085 \\
  & PMF & 0.0883 & 0.0879 & 0.7095 & 0.6791 & 0.0642 & 0.0495 & 0.6669 & 0.6939 & 0.0278 & 0.0230 & 0.7193 & 0.5877 \\
  & DPP & 0.0921 & 0.0979 & 0.7021 & 0.6845 & 0.0721 & 0.0613 & 0.6736 & 0.6967 & 0.0277 & 0.0248 & 0.7444 & 0.6164 \\
  & CATE & 0.0928 & 0.1008 & \underline{0.7599} & \underline{0.7224} & 0.0703 & 0.0606 & \textbf{0.7435} & \underline{0.7406} & 0.0264 & 0.0246 & \underline{0.7676} & \underline{0.6271} \\
  \cmidrule{2-14}
  & D3Rec & \textbf{0.0984} & \textbf{0.1081} & \textbf{0.7838} & \textbf{0.7341} & \textbf{0.0816} & \textbf{0.0735} & \underline{0.7427} & \textbf{0.8000} & \textbf{0.0401} & \textbf{0.0369} & \textbf{0.7795} & \textbf{0.6500} \\
  & \textit{Improv} (\%) & +2.95 & +1.84 & +3.15 & +1.62 & +11.16 & +14.83 & -0.10 & +8.02 & +28.37 & +22.17 & +1.56 & +3.64 \\
 \bottomrule
 \end{tabular}}
 \label{table:main}
 \vspace{-0.2cm}
\end{table*}

\vspace{+0.1cm}
\noindent \textbf{Evaluation Metrics.}
We evaluate the accuracy and diversity with the following metrics over the top-$K$ items, where $K \in \{10,20\}$.
For accuracy, we adopt widely used metrics: Recall$@K$ and NDCG$@K$.
For diversity, we adopt Coverage and Entropy, as done in \cite{DGCN, DPGNN}.
We denote the top-$K$ recommended list for a user $u$ as $\hat{\boldsymbol{x}}_\text{topK} \in \{0, 1\}^{|\mathcal{I}|}$, where $||\hat{\boldsymbol{x}}_\text{topK}||_1 = K$.
Then, the category distribution of the recommendation is obtained as $\hat{\boldsymbol{y}}_\text{topK} = F^\top \hat{\boldsymbol{x}}_\text{topK} \ / \ ||F^\top \hat{\boldsymbol{x}}_\text{topK}||_1$.
Then, the diversity metrics are computed as follows:
\begin{align}
 \text{Coverage}@K & = |\mathcal{U}|^{-1} \cdot \sum_u \frac{\text{non-zero}(\hat{\boldsymbol{y}}_\text{topK})}{|\mathcal{C}|}, \nonumber \\
 \text{Entropy}@K & = |\mathcal{U}|^{-1} \cdot \sum_u \frac{\mathbb{H}[\hat{\boldsymbol{y}}_\text{topK}]}{\text{log}|\mathcal{C}|}.
\end{align}
$\text{non-zero}(\cdot)$ counts the number of non-zero elements and $\mathbb{H}[\cdot]$ denotes the entropy.
Both diversity metrics are bounded in [0,1].

\vspace{+0.1cm}
\noindent \textbf{Methods compared.}
We compare D3Rec with conventional recommenders that consider only accuracy:
\begin{itemize}[leftmargin=*]
    \item \textbf{MultVAE} \cite{multvae} utilizes variational autoencoder to predict user interactions, assuming that the interactions follow a multinomial distribution. 
    \item \textbf{DiffRec} \cite{DiffRec} leverages diffusion framework to predict complex distributions user preferences.
\end{itemize}
and disentangled recommenders to enhance diversity:
\begin{itemize}[leftmargin=*]
    \item \textbf{MacridVAE} \cite{macrid} is a disentangled variational autoencoder that captures users' multiple interests, which is beneficial for diversity, as explored in \cite{DPCML, ComiRec, multi}.
    \item \textbf{DCRS} \cite{DCRS} utilizes disentanglement to capture user preferences for item categories, thereby mitigating the bias towards popular categories.
\end{itemize}
and state-of-the-art methods controlling the accuracy-diversity trade-off:
\begin{itemize}[leftmargin=*]
    \item \textbf{COR} \cite{COR} adopts counterfactual inference for out-of-distribution generalization. We treat the user's category preferences as a changeable user attribute.
    \item \textbf{Dual Process} \cite{DPGNN} proposes a model that integrates the arousal theory of human interest.
    \item \textbf{MMR} \cite{MMR} is a post-processing method to maximize marginal relevance to control the balance between accuracy and diversity. 
    \item \textbf{PMF} \cite{pmf} is a post-processing method that considers the coverage of the user's interests in addition to relevance and diversity.
    \item \textbf{DPP} \cite{dpp} is a post-processing method that balances relevance and diversity using the determinantal point process (DPP).
    \item \textbf{CATE} is a post-processing method proposed in ComiRec \cite{ComiRec}, which balances accuracy and diversity based on the category distribution of the recommendation list.
\end{itemize}

\begin{figure*}[t]
  \centering
  \includegraphics[width=\linewidth]{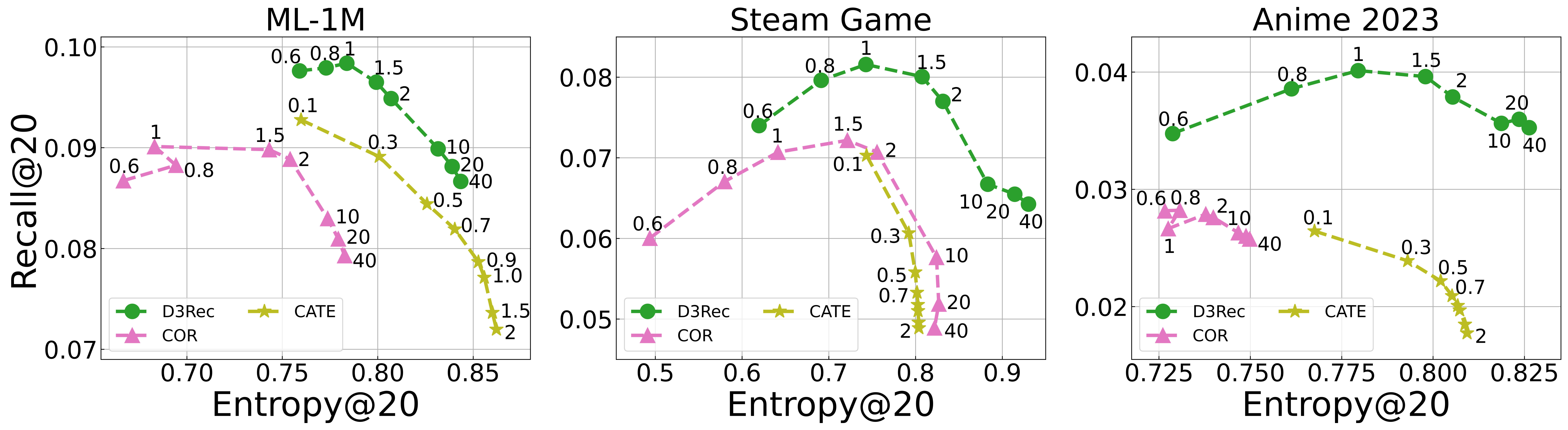}
  % \centering
  % \begin{subfigure}[t]{\linewidth}
  %   \centering
  %   \includegraphics[width=\linewidth]{figure/control.jpg}
  %   \caption{Accuracy and Diversity at different temperatures}
  % \end{subfigure}
  % \begin{subfigure}[t]{\linewidth}
  %   \centering
  %   \includegraphics[width=\linewidth]{figure/curve.jpg}
  %   \caption{Accuracy-Diversity curves}
  % \end{subfigure}
  \vspace{+0.01cm}
  \caption{Accuracy-diversity curves on three real-world datasets. The closer to the top right corner, the better the trade-off between accuracy and diversity.}
  \label{fig:control}
  \vspace{-0.2cm}
\end{figure*}

\vspace{+0.1cm}
\noindent \textbf{Implementation.} \label{sec:hyper}
For each dataset, the best hyperparameters are selected through grid searches on the validation set with early stopping.
We use the AdamW optimizer \cite{adamw} with a learning rate in $\{1e^{-3}, 5e^{-4}, 1e^{-4}, 5e^{-5}\}$ and weight decay in $\{0, 1e^{-1}, 1e^{-2}, 1e^{-3}\}$.
We set the hidden size to [200, 600] and batch size to 400, as done in \cite{DiffRec}.
The dropout ratio is searched from $\{0.1, 0.3, 0.5\}$ for all models.
Model-specific hyper-parameters are searched as follows:
\begin{itemize}[leftmargin=*]
    \item \textbf{MultVAE}, \textbf{MacridVAE}: The regularization strength $\beta$ and the annealing step are chosen from $\{0.2, 0.5, 0.7\}$ and $\{500, 1000, 2000\}$, respectively.
    \item \textbf{Dual Process}: We tune $\alpha$ and $\beta$ in $\{0, 1, 1e^{-1}, 1e^{-2}, 1e^{-3}, 1e^{-4} \}$, which determine the strength of curiosity in the training and inference phases, respectively.
    \item \textbf{Post-processing methods}: 200 items are retrieved using the MultVAE, and re-ranked by MMR, DPP, and CATE. The trade-off parameter is searched in $\{0.1, 0.3, 0.5, 0.7, 0.9\}$. For PMF, the control parameters $\alpha$ and $\beta$ are chosen from $\{0.1, 0.2, 0.3, 0.4, 0.5\}$.
    \item \textbf{DiffRec}, \textbf{D3Rec}: We search the diffusion step $T$ from $\{10, 15, 20, \\ 100\}$. The noise scale, noise lower bound, and noise upper bound are searched in $\{1, 1e^{-2}, 1e^{-4}\}$, $\{5e^{-3}, 1e^{-4}, 5e^{-4}\}$, and $\{5e^{-2}, \\ 1e^{-3}, 5e^{-3}\}$, respectively. Additionally, for D3Rec, we choose $\lambda$ from $\{1, 1e^{-2}, 1e^{-4}\}$, the guiding strength $w$ from $\{-0.7, -0.5, -0.3, \\ 0, 0.3, 0.5\}$, loss weight upper bound $\gamma_\text{min}$ from \{0.3, 0.5, 0.8, 1\}, and lower bound $\gamma_\text{max}$ from \{1, 1.3, 1.6, 2\}.
\end{itemize}
\vspace{-0.3cm}

\subsection{Performance Evaluation}
We evaluate the performance of D3Rec and the compared methods from the perspectives of accuracy and diversity. 
For each method, the trade-off hyper-parameters are selected based on the highest harmonic mean of Recall$@20$ and Entropy$@20$.
Table \ref{table:main} presents the average performance of five different random seeds.
For all metrics, except Entropy@20 of the Steam Game dataset, the improvements over the best baseline are statistically significant (p-value < 0.05) under one-sample t-tests.
Specifically, we found following observations.

\vspace{+0.1cm}
\noindent \textbf{Advantages of using category information.} 
D3Rec, COR, DCRS, and Dual Process, which leverage category information, outperform MacriVAE and post-processing methods in terms of both accuracy and diversity.
% Although the improvement of COR, DCRS and Dual Process is slightly low due to the outdated information in the past interactions, 
This finding indicates that category data are crucial in resolving the accuracy-diversity dilemma.
Especially, it is important when considering the positions in the ranked list, as the NDCG values are higher than those of MultVAE across all datasets.

\vspace{+0.1cm}
\noindent \textbf{Effectiveness of diffusion process.}
D3Rec and DiffRec, which utilize the diffusion process, outperform compared methods in both accuracy and diversity metrics on most datasets.
This is because the user interactions contain biased category preferences, and therefore, generated recommendations exhibit the risk of bias amplification.
On the other hand, diffusion-based recommenders first corrupt the user interactions and effectively eliminate the category preferences lurking in user interactions.
% \item D3Rec outperforms all compared methods. It mitigates the impact of outdated information by corrupting user's interactions and generates more diverse items based on category preferences.
% Orienting toward category preference addresses the accuracy-diversity dilemma because it increases diversity while mitigating bias amplification.

\subsection{Controlling Diversity with Temperature}
To demonstrate the effectiveness of D3Rec in achieving a better trade-off between accuracy and diversity, we visualize Pareto curves \cite{pareto} on three real-world datasets.
We adjust the diversity based on users' original category preferences using temperature: $\tilde{\boldsymbol{y}} = \text{Softmax}(\text{log}(\boldsymbol{y}) / \tau)$.
% Specifically, we tune the $\tau$ in $\{0.1, 0.3, 0.5, 0.8, 1, 1.2, 1.5, 2, 5, 10, 20\}$.
% Figure \ref{fig:control} presents the Recall-Entropy curves for D3Rec, COR \cite{COR}, which is the best end-to-end approach in our experiment, and CATE \cite{ComiRec}, which is the state-of-the-art post-processing approach.
Figure \ref{fig:control} presents the Recall-Entropy curves for D3Rec, COR \cite{COR}, and CATE \cite{ComiRec}.
COR and CATE are the best-performing end-to-end and post-processing approaches with controllability in our experiment, respectively.
We emphasize the following two key strengths of D3Rec.

\vspace{+0.1cm}
\noindent \textbf{Monotonic diversity control.} 
The large temperature $\tau$ exhibits a smoother targeted category distribution $\tilde{\boldsymbol{y}}$, and therefore, needs to be associated with a more diverse recommendation.
We observe that D3Rec is able to control the diversity monotonically according to $\tau$, while COR yields a more diverse recommendation with a smaller temperature ($\tau=0.8$ has a larger Entropy@20 than $\tau=1$ in ML-1m).
On the other hand, although CATE can monotonically control diversity, its effect is minimal, especially on the Anime 2023 dataset.
This result implies that D3Rec effectively captures the desired level of diversity with our category preference representation and UNet-like encoder-decoder framework.

\vspace{+0.1cm}
\noindent \textbf{Pareto frontier in accuracy-diversity trade-off.}
For Pareto curves in Figure \ref{fig:control}, the closer to the top right corner, the better the trade-off between accuracy and diversity.
We observe that D3Rec is the closest to the top-right corner, i.e., the Pareto frontier.
This indicates that, given the equal accuracy, D3Rec achieves better diversity than others, and vice versa.
Moreover, D3Rec exhibits a smaller performance drop in Recall@20 when adjusting the diversity (i.e., Entropy@20). 
% At the same time, it suggests that improving diversity conditioned on users' category preferences is beneficial to addressing the accuracy-diversity dilemma.
Similar results are observed for other metrics, such as NDCG@20 and Coverage@20.

\begin{table}[t]
  \caption{Synthetic Data statistics}
  \vspace{+0.05cm}
   \renewcommand{\arraystretch}{0.6}
  \begin{tabular}{cccccc}
    \toprule
    Dataset & \#Train & \#Valid & \#Test & $C_{KL}$ \\
    \midrule
    ML-1M & 392,990 & 100,842 & 59,445 & 2.01 \\
    Steam Game & 901,764 & 241,148 & 208,747 & 1.64 \\
    Anime 2023 & 8,736,817 & 2,256,163 & 1,810,009 & 1.63 \\
    \bottomrule
  \end{tabular}
  \label{table:syn_statistics}
  % \vspace{-0.3cm}
\end{table}

\subsection{Controlling Diversity with Arbitrary Targeted Category Preferences}
We show the superiority of D3Rec in fast adaptation to arbitrary targeted category preferences that significantly differ from the original category preferences.

\vspace{+0.1cm}
\noindent \textbf{Semi-synthetic datasets.}
We construct three semi-synthetic datasets based on the real-world datasets, as follows:
\begin{enumerate}[leftmargin=*]
    \item For each user, we select the bottom 30\% categories among those the user has consumed.
    \item We consider items belonging to those bottom 30\% categories as test interactions $\boldsymbol{x}_{\text{test}} \in \{0,1\}^{|\mathcal{I}|}$.
    \item Excluding the test interactions, we sort the remaining data chronologically and split it into training interactions (80\%) and validation interactions (20\%).
    % \item For the users whose interactions cannot be split into three parts, we drop them. 
    % Specifically, there is one such user in the Anime 2023 dataset.
\end{enumerate}
% \noindent For example, let's consider a user who showed preferences of $[0.3, 0.25, 0.25, 0.15, 0.05]$ for five categories.
% Then, items belonging to the fourth and the fifth categories are set as the test interactions.
The test interactions exhibit category preferences that significantly differ from the training interactions, and therefore, are suitable to examine the adaptation ability of D3Rec.
We compare the Kullback-Leibler divergence of category distribution between the training and the test interactions \cite{Calib}.
A higher $C_\text{KL}$ indicates a greater mismatch in category preferences between the training dataset and the test dataset.
The data statistics for three semi-synthetic datasets are presented in Table \ref{table:syn_statistics}.

\vspace{+0.1cm}
\noindent \textbf{Adaptation to arbitrary targeted category preferences.}
We train MultVAE, COR, and D3Rec on the training interactions, and inference with the category distribution of the test interactions, i.e., $\tilde{\boldsymbol{y}}_{\text{test}} = \boldsymbol{F}^\top  \boldsymbol{x}_\text{test} \ / \ ||\boldsymbol{F}^\top  \boldsymbol{x}_\text{test}||_1$.
We tune hyper-parameters according to Section \ref{sec:hyper} and fix the guiding strength $w$ at 7 for D3Rec. 
We adopt Recall@K and NDCG@K to compare strong generalization and fast adaptation.
Table \ref{table:syn_perform} demonstrates that D3Rec outperforms MultVAE and COR, demonstrating its fast adaptation to the arbitrary targeted category preferences.
Notably, we observe that the fast adaptation of D3Rec is more effective on datasets with higher $C_\text{KL}$, i.e., ML-1M.
\begin{table}[t]
 \caption{Performance of adaptation ability to arbitrary targeted category preferences on three synthetic datasets.}
 \vspace{+0.1cm}
 \resizebox{\linewidth}{!}{
 \begin{tabular}{c| c | cccc }
 \toprule
 Dataset & Model & Recall@10 & Recall@20 & NDCG@10 & NDCG@20 \\
\midrule
\midrule
 \multirow{4}{*}{ML-1M} & MultVAE & 0.0257 & 0.0523 & 0.0202 & 0.0316 \\
 & COR & \underline{0.1046} & \underline{0.1736} & \underline{0.1126} & \underline{0.1363} \\
 & D3Rec & \textbf{0.2542} & \textbf{0.3735} & \textbf{0.3134} & \textbf{0.3403} \\
 & \textit{Improv} (\%) & +143.05 & +115.23 & +178.36 & +149.75 \\
 \midrule
 \multirow{4}{*}{Steam Game} & MultVAE & 0.0077 & 0.0143 & 0.0065 & 0.0091 \\
 & COR & \underline{0.1437} & \underline{0.2008} & \underline{0.1444} & \underline{0.1615} \\
 & D3Rec & \textbf{0.1606} & \textbf{0.2331} & \textbf{0.1467} & \textbf{0.1703} \\
 & \textit{Improv} (\%) & +11.75 & 16.12 & +1.60 & +5.47 \\
\midrule
\multirow{4}{*}{Anime 2023} &  MultVAE & 0.0100 & 0.0170 & 0.0099 & 0.00129 \\
 & COR & \underline{0.2204} & \underline{0.3116} & \underline{0.3093} & \underline{0.3205} \\
 & D3Rec & \textbf{0.2314} & \textbf{0.3440} & \textbf{0.3233} & \textbf{0.3437}  \\
 & \textit{Improv} (\%) & +4.99 & +10.39 & +4.53 & +7.27 \\
 \bottomrule
 \end{tabular}}
 \label{table:syn_perform}
 % \vspace{-0.3cm}
\end{table}

% &  & 
% &  & 
 % &  & 

% &  & 

 % & \multicolumn{4}{c|}{Steam Game} & \multicolumn{4}{c}{Anime 2023} \\

\subsection{In-depth Analyses}

\begin{figure}[t]
  \centering
  \includegraphics[width=\linewidth]{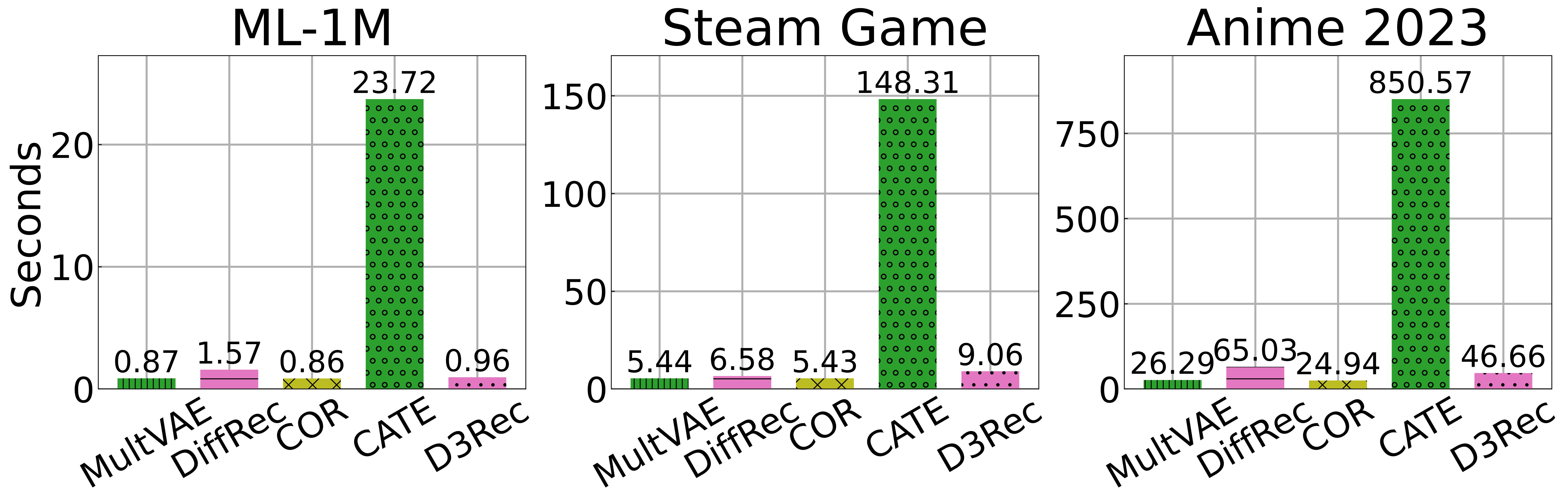}
  \vspace{+0.03cm}
  \caption{Inference time comparison.}
  \label{fig:inference_time}
  % \vspace{-0.5cm}
\end{figure}

\vspace{+0.1cm}
\noindent \textbf{Inference burden.}
Figure \ref{fig:inference_time} shows the inference time of D3Rec and baselines on three real-world datasets.
MultVAE and DiffRec do not consider diversity and exhibit the lowest inference burden.
% The end-to-end approach (i.e., COR) decides the accuracy-diversity trade-off during training, therefore, the inference time is almost identical to MultVAE, which is the base architecture of COR.
The end-to-end approach (i.e., COR) does not require any additional post-processing, therefore, the inference time is almost identical to MultVAE, which is the base architecture of COR.
The post-processing approach (i.e., CATE) needs an additional process after the inference, and therefore, takes the largest inference latency.
Lastly, D3Rec exhibits an inference burden similar to DiffRec, since their generation process takes similar computational effort.
It is noted that DiffRec generally requires more denoising steps ($T$) than D3Rec for the best performance.

\vspace{+0.1cm}
\noindent \textbf{Robustness on the noisy dataset.}
\begin{table}[t]
\caption{Performance comparison on clean and noisy Steam Game.}
 \vspace{+0.1cm}
 \resizebox{\linewidth}{!}{
\begin{tabular}{c|cc|cc|cc}
\toprule
\multirow{2}{*}{Dataset} & \multicolumn{2}{c|}{MultVAE} & \multicolumn{2}{c|}{DiffRec} & \multicolumn{2}{c}{D3Rec} \\
& Recall@20 & NDCG@20 & Recall@20 & NDCG@20 & Recall@20 & NDCG@20 \\
\midrule
\midrule
Clean & 0.0724 & 0.0617 & 0.0710 & 0.0605 & 0.0816 & 0.0735 \\
Noisy & 0.0702 & 0.0584 & 0.0677 & 0.0578 & 0.0806 & 0.0734 \\
\midrule
Change (\%) & -3.04\% & -5.65\% & -4.87\% & -4.67\% & \textbf{-1.23\%} & \textbf{-0.14\%} \\
\bottomrule
\end{tabular}}
\label{tab:exp_noise}
\end{table}
Table \ref{tab:exp_noise} highlights the comparative performance of three models, including MultVAE, DiffRec, and D3Rec, on both clean and noisy versions of the Steam Game dataset.
When noise is introduced into the training data (with a ratio of 0.3), all models show a decrease in performance.
Meanwhile, D3Rec demonstrates remarkable robustness, with only a minor drop of -1.23\% in Recall@20 and -0.14\% in NDCG@20.
This minimal degradation contrasts sharply with the more significant declines observed in MultVAE and DiffRec, which show drops of -3\% $\sim$ -5.6\%.
The results underscore the effectiveness of D3Rec in maintaining performance under noisy training conditions, suggesting that it is better equipped to handle data perturbations compared to MultVAE and DiffRec.

\vspace{+0.1cm}
\noindent \textbf{Ablation study.}
We analyze the impact of each component in D3Rec: two-tower encoders (two encoder), 
$\mathcal{L}_\text{ortho}$, $\mathcal{L}_\text{cate}$, $\mathcal{L}_\text{emb}$, and the re-weight strategy (re-weight).
We conduct ablation experiments by removing each component.
Table \ref{table:ablation} shows the ablation study of D3Rec on Steam Game and semi-synthetic Steam Game.
The two-tower encoders and the orthogonal regularization successfully disentangle the category features and guide D3Rec to only manipulate the category preferences, resulting in a better accuracy-diversity trade-off.
This verifies the effectiveness of disentanglement for controllability as it explicitly encodes category-aware representation.
We observe that the re-weight strategy significantly impacts diversity because it addresses the imbalance issue of category preferences.
% Similar results are observed for other datasets, such as ML-1M and Anime.

\begin{table}[t]
 \caption{
 Ablation study of D3Rec on Steam Game and synthetic Steam Game datasets.
 }
 \vspace{+0.1cm}
 \resizebox{\linewidth}{!}{
 \begin{tabular}{c | cc | cc }
 \toprule
\multirow{2}{*}{Method} & \multicolumn{2}{c|}{Steam Game} & \multicolumn{2}{c}{Synthetic Steam Game} \\
& Recall@20 & Entropy@20 & Recall@20  & NDCG@20 \\
\midrule
\midrule
 w/o two encoder & \underline{0.0813} & 0.\textbf{7496} & 0.2286 & 0.1612 \\
 w/o $\mathcal{L}_\text{ortho}$ & 0.0800 & \underline{0.7435} & 0.2284 & 0.1607 \\
 w/o $\mathcal{L}_\text{cate}$ & 0.0798 & 0.7236 & 0.2293 & 0.1680\\
 w/o $\mathcal{L}_\text{emb} $ & 0.0806 & 0.7310 & 0.2306 & \underline{0.1691} \\
 w/o re-weight & 0.0809 & 0.7029 & \underline{0.2320} & 0.1671 \\
 \midrule
 D3Rec & \textbf{0.0816} & 0.7427 & \textbf{0.2331} & \textbf{0.1703} \\
 \bottomrule
 \end{tabular}}
 \label{table:ablation}
 % \vspace{-0.2cm}
\end{table}

\begin{figure}[t]
  \centering
  \includegraphics[width=\linewidth]{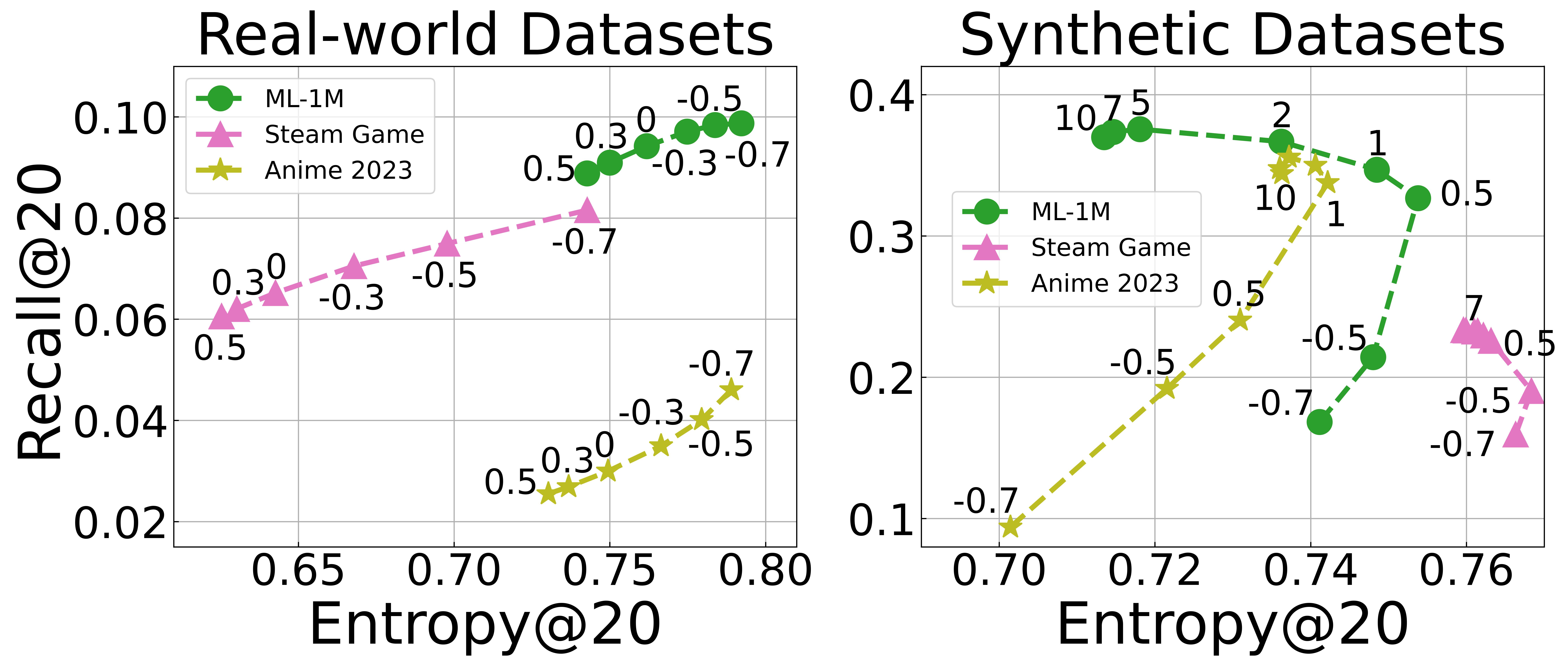}
  \vspace{+0.1cm}
  \caption{Effect of the guiding strength $w$ on real-world (left) and synthetic datasets (right).}
  \label{fig:control_w}
  % \vspace{-0.5cm}
\end{figure}

% \vspace{-0.5cm}

\vspace{+0.1cm}
\noindent \textbf{Effect of the guiding strength.}
We adjust the value of $w$ in Eq.\ref{eq:pred_conditioned} to explore the effect of guiding category preferences.
In Figure \ref{fig:control_w} left, we find that excessive reliance on category preferences (i.e., large $w$) can lead to decreased performance on the real-world datasets.
This occurs because users consider various factors beyond just categories.
Overemphasizing category preferences alone, without accounting for these additional factors, can harm overall performance.
It is noted that we disregard small $w$ values (i.e., $w < -0.7$) as they lose controllability.
On the other hand, in Figure \ref{fig:control_w} right, strong guidance $w > 0.5$ exhibits superiority.
Synthetic test sets have significantly different category distributions from the training sets, and therefore, strong guidance for the target category preferences is required.

\vspace{+0.1cm}
\noindent \textbf{Effect of the diffusion steps.}
% First Table
\begin{table}[t]
\caption{Effect of the diffusion steps $T$ on Steam dataset.}
 \vspace{+0.1cm}
 \resizebox{\linewidth}{!}{
\begin{tabular}{c|cccc}
\toprule
$T$ & Recall@20 & NDCG@20 & Entropy@20 & Coverage@20 \\
\midrule
\midrule
10  & 0.078 & 0.068 & 0.742 & 0.802 \\
15  & \textbf{0.082} & \textbf{0.073} & 0.743 & 0.800 \\
20  & 0.052 & 0.050 & \textbf{0.794} & 0.845 \\
100 & 0.045 & 0.044 & 0.793 & \textbf{0.865} \\
\bottomrule
\end{tabular}}
\label{tab:hyper_step}
\end{table}
Table \ref{tab:hyper_step} investigates the impact of varying the diffusion step $T$ on the performance of D3Rec on the Steam dataset.
Similar results were observed for ML-1M and Anime datasets.
As the steps increase, the influence of guidance also grows, leading to lower accuracy and greater diversity.
A larger diffusion step can enhance the model's ability to recommend a broader range of items, however, overly high diffusion steps negatively impact the model’s ranking capabilities.
This analysis suggests that choosing the appropriate $T$ depends on whether the application prioritizes ranking accuracy or a more diverse set of recommendations.
It is noted that DiffRec generally requires more denoising steps ($T \approx 100$) than D3Rec for the best performance.
We also analyzed the impact of the inference step $T’$ and found no significant differences.

% \vspace{+0.1cm}
% \noindent \textbf{Effect of the noise scale.}
% \input{tables/5exp_hyper_noise}
% In this work, we adopt a linear decay for noise scheduling introduced in DiffRec \cite{DiffRec}, as follows:
% \begin{equation}
% 1 - \bar{\alpha}_t = s \cdot \left[ \alpha_{\text{min}} + \frac{t - 1}{T - 1} (\alpha_{\text{max}} - \alpha_{\text{min}}) \right].
% \end{equation}
% Table \ref{tab:hyper_noise} explores the influence of the noise scale $s$ on the Steam dataset.
% We observe that the linear 
% Similar results were observed for ML-1M and Anime datasets.

\section{Conclusion}
We claim that existing diverse recommender systems lack controllability at inference, and devise three desiderata for controlling diversity with category preferences.
Based on our desiderata, we propose D3Rec, an end-to-end method that controls the accuracy-diversity trade-off at the inference phase.
In the forward process, D3Rec eliminates category preferences lurking within user interactions by adding noises.
Then, in the reverse process, D3Rec generates recommendations through denoising steps while reflecting the targeted category preference.
Moreover, D3Rec can adapt to arbitrary category preferences that deviate from the original user category preferences.
Our extensive experiments on both real-world and synthetic datasets have demonstrated the superiority and in-depth analyses of D3Rec.

\section*{Acknowledgement}
This work was supported by the IITP grant funded by the MSIT (South Korea, No.2018-0-00584, No.2019-0-01906), the NRF grant funded by the MSIT (South Korea, No.RS-2023-00217286, No. RS-2024-00335873), and NRF grant funded by MOE (South Korea, 2022R1\\A6A1A03052954).

%%
%% The next two lines define the bibliography style to be used, and
%% the bibliography file.
\bibliographystyle{ACM-Reference-Format}
\bibliography{sample-base}

%%
%% If your work has an appendix, this is the place to put it.
\clearpage % move to another page
\begin{appendices}
\nobalance
\section{Detailed Experimental Setup}
\label{sec:detailexp}
\vspace{+0.1cm}
\noindent \textbf{Datasets.}
We adopt three datasets in different domains.
\begin{itemize}[leftmargin=*]
    \item \textbf{ML-1M}\footnote{https://grouplens.org/datasets/movielens/}: This dataset comprises user ratings for movies ranging from 1 to 5. We discard user interactions with ratings less than 4, setting $\boldsymbol{x}_i = 1$ if a user rates an item $i$ as 4 or higher, and $\boldsymbol{x}_i = 0$ otherwise. Subsequently, we sort all interactions chronologically based on the timestamps.
    \item \textbf{Steam Game} \cite{steam}: This dataset contains user reviews from the Steam video game platform. We set $\boldsymbol{x}_i = 1$ if a user has reviewed item $i$, and $\boldsymbol{x}_i = 0$ otherwise. We, then, sort all interactions chronologically based on the timestamps.
    \item \textbf{Anime 2023}\footnote{https://www.kaggle.com/datasets/dbdmobile/myanimelist-dataset/data}: This dataset consists of user ratings for anime on a scale of 1 to 10. We set $\boldsymbol{x}_i = 1$ if a user rates an item $i$ as 8 or higher, and $\boldsymbol{x}_i = 0$ otherwise. As this dataset does not include timestamps, we randomly shuffle the user interactions.
\end{itemize}
\noindent For all datasets, we consider an item's genres as the categories.
We adopt 20-core settings for users, items, and categories across all datasets to ensure data quality \cite{DCRS}.
After that, we split each user’s interactions into a training set (60\%), a validation set (20\%), and a test set (20\%).
The statistics of the three datasets after the pre-processing are presented in Table \ref{table:statistics}.

\vspace{+0.1cm}
\noindent \textbf{Methods compared.}
We compare D3Rec with conventional recommenders that consider only accuracy:
\begin{itemize}[leftmargin=*]
    \item \textbf{MultVAE} \cite{multvae} utilizes variational autoencoder to predict user interactions, assuming that the interactions follow a multinomial distribution. 
    \item \textbf{DiffRec} \cite{DiffRec} leverages diffusion framework to predict user interactions. Diffusion enables it to model complex distributions.
\end{itemize}
, disentangled recommenders to promote diversity:
\begin{itemize}[leftmargin=*]
    \item \textbf{MacridVAE} \cite{macrid} is a disentangled variational autoencoder that learns disentangled representations from user behavior. Inspired by that capturing users' multiple interests is beneficial for diversity, as explored in \cite{DPCML, ComiRec, multi}, we learn multiple disentangled user representations that capture preferences over each category.
    \item \textbf{DCRS} \cite{DCRS} utilizes disentanglement to capture user preferences for item categories, thereby mitigating the accuracy-diversity dilemma. We employ MultVAE as the base model and disentangle the encoder's output into representations that are dependent and independent of category preference distribution.
\end{itemize}
and state-of-the-art methods controlling the accuracy-diversity trade-off:
\begin{itemize}[leftmargin=*]
    \item \textbf{COR} \cite{COR} adopts counterfactual inference for strong out-of-distribution generalization and fast adaptation. We treat the user's category preferences as a changeable user attribute.
    \item \textbf{Dual Process} \cite{DPGNN} proposes a model that integrates the arousal theory of human interest. While using MultVAE as the base model, to reflect dual processing, we treat the output of the encoder as the user's embedding and regard the decoder as a network updated through the heuristic channel.
    \item \textbf{MMR} \cite{MMR} is a traditional re-ranking model designed to maximize marginal relevance for controlling the balance between accuracy and diversity. 
    \item \textbf{PMF} \cite{pmf} is a re-ranking model that further considers the coverage of the user's interests in addition to relevance and diversity.
    \item \textbf{DPP} \cite{dpp} is a re-ranking model that balances relevance and diversity using the determinantal point process (DPP). Through greedy maximum a posteriori inference, it reduces time complexity.
    \item \textbf{CATE} is a re-ranking model proposed in ComiRec \cite{ComiRec}, which balances accuracy and diversity based on the category distribution of the recommended list.
\end{itemize}

\vspace{+0.1cm}
\noindent \textbf{Hyper-parameter settings.}
For each dataset, the best hyperparameters are selected through grid searches on the validation set with early stopping. We tune the AdamW optimizer \cite{adamw} with a learning rate in $\{1e^{-3}, 5e^{-4}, 1e^{-4}, 5e^{-5}\}$ and weight decay in $\{0, 1e^{-1}, 1e^{-2}, 1e^{-3}\}$. We set the hidden size to [200, 600], batch size to 400, like \cite{DiffRec}, and the dropout ratio is searched from $\{0.1, 0.3, 0.5\}$ for all models. Model-specific hyper-parameters are searched as follows.
\begin{itemize}[leftmargin=*]
    \item \textbf{MultVAE} and \textbf{MacridVAE}: The regularization strength $\beta$ and the annealing step are chosen from $\{0.2, 0.5, 0.7\}$ and $\{500, 1000, 2000\}$.
    \item \textbf{Dual Process}: We tune $\alpha$ and $\beta$ in $\{0, 1, 1e^{-1}, 1e^{-2}, 1e^{-3}, 1e^{-4} \}$, which determine the strength of curiosity in the training and inference phases, respectively.
    \item \textbf{DiffRec} and \textbf{D3Rec}: We search for the total diffusion step $T$ in $\{5, 15, 40, 100\}$ and fix the sampling step $T^\prime$ to 0, as in \cite{DiffRec}. The noise scale, noise lower bound, and noise upper bound are searched in $\{1, 1e^{-2}, 1e^{-4}\}$, $\{5e^{-3}, 1e^{-4}, 5e^{-4}\}$, and $\{5e^{-2}, 1e^{-3}, 5e^{-3}\}$, respectively. Additionally, for D3Rec, the ratio of the condition dropout is searched in $\{0.1, 0.3, 0.5\}$, $\lambda$ in $\{1, 1e^{-2}, 1e^{-4}\}$, the guiding strength $w$ in $\{-0.7, -0.5, -0.3, 0, 0.3, 0.5\}$, loss weight upper bound $\gamma_\text{min}$ in \{0.3, 0.5, 0.8, 1\} and lower bound $\gamma_\text{max}$ in \{1, 1.3, 1.6, 2\}.
    \item \textbf{Post-processing methods}: The top 200 items are retrieved using the MultVAE, and re-ranked by MMR, DPP, and CATE. The trade-off parameter is searched in $\{0.1, 0.3, 0.5, 0.7, 0.9\}$. In PMF, the control parameters $\alpha$ and $\beta$ are chosen from $\{0.1, 0.2, 0.3, 0.4, 0.5\}$.
\end{itemize}
\end{appendices}

\end{document}